\documentclass[a4paper,11pt]{article}

\renewcommand{\thetable}{\Roman{table}}

\pdfoutput=1 % if your are submitting a pdflatex (i.e. if you have
\usepackage{jcappub} % for details on the use of the package, please
\usepackage[T1]{fontenc} % if needed
\usepackage{lineno,hyperref}
\usepackage{caption}
\usepackage{subcaption}
\usepackage{soul}

\title{\boldmath On the use of ferroelectric material in the detection of dark matter axions}

\author[*,a]{J.M.~García Barceló,}
\author[a]{A.~\'Alvarez~Melc\'on,}
\author[b]{S.~Arguedas~Cuendis,}
\author[a]{A.~D\'iaz-Morcillo,}
\author[c]{B.~Gimeno,}
\author[d]{A.~Kanareykin,}
\author[a]{A.J.~Lozano-Guerrero,}
\author[a]{P.~Navarro,}
\author[e]{and W.~Wuensch}

\affiliation[a]{Department of Information and Communications Technologies, Technical University of Cartagena, 30203 - Cartagena, Spain}
\affiliation[b]{Institut de Ciències del Cosmos, Universitat de Barcelona (UB-IEEC), 08028 - Barcelona, Catalonia, Spain}
\affiliation[c]{Instituto de F\'isica Corpuscular (IFIC), CSIC-University of Valencia, 46980 - Valencia, Spain}
\affiliation[d]{Euclid Techlabs, LLC, Bolingbrook, IL}
\affiliation[e]{European Organization for Nuclear Research (CERN), 1211 Geneva 23, Switzerland}

\affiliation[*]{Corresponding author}

\emailAdd{josemaria.garcia@upct.es}

\abstract
{Tuning is an essential requirement for the search of dark matter axions employing haloscopes since its mass is not known yet to the scientific community. At the present day, most haloscope tuning systems are based on mechanical devices which can lead to failures due to the complexity of the environment in which they are used. However, the electronic tuning making use of ferroelectric materials can provide a path that is less vulnerable to mechanical failures and thus complements and expands current tuning systems. In this work, we present and design a novel concept for using the ferroelectric Potassium Tantalate ($KTaO_3$ or KTO) material as a tuning element in haloscopes based on coupled microwave cavities. In this line, the structures used in the Relic Axion Detector Exploratory Setup (RADES) group are based on several cavities that are connected by metallic irises, which act as interresonator coupling elements. In this article, we also show how to use these $KTaO_3$ films as interresonator couplings between cavities, instead of inductive or capacitive metallic windows used in the past. These two concepts represent a crucial upgrade over the current systems employed in the dark matter axions community, achieving a tuning range of $2.23 \, \%$ which represents a major improvement as compared to previous works ($<0.1 \, \%$) for the same class of tuning systems. The theoretical and simulated results shown in this work demonstrate the interest of the novel concepts proposed for the incorporation of this kind of ferroelectric media in multicavity resonant haloscopes in the search for dark matter axions.}

\begin{document}
\maketitle
\flushbottom

\section{Introduction}
\label{Introduction}
In recent years there has been a high interest for the search of axions and other particles compatible with the Standard Model which could be part of dark matter. Axions, particles predicted by Weinberg~\cite{Weinberg:1978} and Wilczek~\cite{Wilczek:1978} could also solve the strong Charge Conjugation-Parity problem \cite{Peccei:1977Jun,Peccei:1977Sep}.\\

Several experimental groups have developed in the last 30 years structures for the detection of such particles \cite{Irastorza:2018dyq} based on the inverse Primakoff effect \cite{Primakoff:1951}. This kind of detectors are divided into several types: light-shining-through wall experiments, which create artificial axions at the laboratory; helioscopes, which search for axions from the Sun (plasma photons in the solar inner layers); and haloscopes, which search for axions in the galactic halo. These systems use the axion-photon conversion effect driven by the action of a powerful external static magnetic field. For the latter, this coupling is enhanced when it occurs in a resonant device like a microwave resonant cavity, as described in \cite{Sikivie:1983ip}.\\

In the last five years, the RADES group has worked with this concept, studying, designing, manufacturing, and taking data with axion detectors (haloscopes) searching for dark matter axions of masses around $\sim 34 \, \mu$eV, although the Ultra High Frequency (UHF) band is now also being investigated for new haloscope designs. The first haloscope developed by RADES, working at $\sim8.4$~GHz, is based on an array of five copper-coated stainless steel cavities cascade connected by four inductive windows (interresonator coupling irises) a concept commonly employed in microwave filters \cite{Cameron}. Another haloscope was designed and manufactured, using an alternating system of irises \cite{RADES_paper2} (inductive and capacitive windows). The size of an individual cavity sets the working frequency (and the axion mass to be explored) since the resonance is determined by the cavity dimensions. The advantage of this kind of haloscopes is that we can increase the volume of the structure without decreasing the working frequency (which is a common problem in the axion community), just by adding more cavities to the structure. However, the larger the haloscope volume, the smaller the mode separation with respect to neighbour resonances (this situation could cause problems in the response if a non-desired mode is very close to the axion one), which in turn can also be controlled by the interresonator coupling. The theoretical foundation of the multicavity behaviour can be found in \cite{RADES_paper1}.\\

The whole axion-search system is based on several components. First, due to the extremely low axion-photon coupling a cryogenic environment (few Kelvin) is needed to decrease the thermal noise. Second, the RADES haloscopes are connected to a receiver which amplifies, filters and down-converts the received radio frequency (RF) power with very low noise levels. And third, the receiver carries out the Analog-Digital conversion and the Fast Fourier Transform for the post-processing of the data taking \cite{RADESreviewUniverse}.\\

The main objectives for an efficient axion detection system are to maximize the power converted from the axion-photon coupling, to increase the analysed axion mass range (and its scanning frequency rate) as well as to optimize the haloscope sensitivity. The RF power detected ($P_d$) depends on properties intrinsic to the axion and on the experimental cavity parameters, namely \cite{RADESreviewUniverse}:
\begin{equation}
\label{Pd}
    P_d \, = \, \kappa \, g^2_{a\gamma}\, \frac{\rho_a}{m_a} \, B_e^2 \, C \, V \, Q_l,
\end{equation}
where $\kappa$ is the coupling to the external receiver  ($\kappa=0.5$ for critical coupling operation regime), $g_{a\gamma}$ is the unknown axion-photon coupling, $\rho_a$ is the dark matter density, $m_a$ is the axion mass, $B_e$ is the external static magnetic field (depends on the magnet used for the experiment), $C$ is the form factor, $V$ is the volume of the cavity and $Q_l$ its loaded quality factor. The unloaded quality factor $Q_0$ is also commonly used in the characterization of resonant cavities since, contrary to $Q_l$, it is independent on the external coupling \cite{pozar}. The form factor ($C$), which measures the coupling between the external magnetostatic field and the RF electric field induced by the axion-photon conversion can be expressed as:
\begin{equation}
\label{eq:C}
    C \, = \, \frac{|\int _V \, \vec{E} \cdot \vec{B}_e \, dV|^2}{\int_V \, |\vec{B}_e|^2 \, dV \int_V \, \varepsilon_r \, |\vec{E}|^2 \, dV},
\end{equation}
where $\varepsilon_r$ is the relative electrical permittivity within the cavity of volume $V$. A measure of the sensitivity of the haloscope is the axion-photon coupling that can be detected for a given signal to noise ratio ($\frac{S}{N}$), which can be obtained by
\begin{equation}
\label{eq:ga}
    g_{a\gamma} \, = \,  \left(\frac{\frac{S}{N} \, k_B \, T_{sys}}{\kappa \, \rho_a \, C \, V \, Q_l}\right)^{\frac{1}{2}}\frac{1}{B_e}\left(\frac{m_a^3}{Q_a \, \Delta t}\right)^{\frac{1}{4}},
\end{equation}
where $k_B$ is the Boltzmann constant, $T_{sys}$ the noise temperature of the system, $\Delta t$ the data taking time window, and $Q_a$ the quality factor of the axion resonance \cite{RADESreviewUniverse}. In summary, the parameters that can be controlled in the haloscope design are $\kappa$, $C$, $V$ and $Q_l$.\\

The frequency tuning in a haloscope is an extremely important feature because the axion mass is unknown. The data taking will be based on scanning a specific mass range of the whole spectrum, so a frequency shifting procedure will be needed. This work is focused on a system for improving the tuning in order to be able to easily scan a specific axion mass region using ferroelectric media. In addition, the scanning rate $\frac{dm_a}{dt}$ is generally used to measure the performance of a haloscope, which can be obtained from equation~\ref{eq:ga} \cite{RADESreviewUniverse}:
\begin{equation}
\label{eq:dmadt}
    \frac{dm_a}{dt} \, = \, Q_a \, Q_l \, \kappa^2 \, g_{a\gamma}^4 \, \left(\frac{\rho_a}{m_a} \right)^2 \, B_e^4 \, C^2 \, V^2 \, \left(\frac{S}{N} \, k_B \, T_{sys}\right)^{-2}.
\end{equation}\\

So far, all the tuning mechanisms employed by haloscope experiments are based on mechanical systems. The ADMX \cite{Stern:2015,Boutan:2018}, HAYSTAC \cite{Zhong:2018} and IBS/CAPP \cite{Choi:2021} collaborations use cylindrical cavities with one or more rods. The tuning is accomplished by rotating these metallic rods inside the cavities, which affect to the electromagnetic field distribution of the resonance inside the cavity thus modifying the resonant frequency of the desired mode. The rotational movement of the rod system is obtained by a series of gears connected to a driven motor or by means of piezoelectric actuators. In the case of the QUAX experiment \cite{Alesini:2020}, movable sapphire shells are used to change the resonant frequency of the cavity. On the other hand, the CAST-CAPP/IBS group uses two movable dielectric sapphire plates placed in parallel and symmetrically at the cavity sides \cite{Miceli:2015}. Also the RADES group has studied a mechanical tuning mechanism based on splitting the haloscope in two identical halves which can be moved symmetrically to increase the effective width dimension of the cavities, thus modifying the axion frequency search range \cite{RadesProceeding:2020}.\\

In 2018, Euclid Techlabs \cite{EuclidTechlabs} proposed to implement a KTO ferroelectric tuning system for dark matter axion searches in the RADES \cite{RADES_paper1,RADESreviewUniverse} and ADMX \cite{Braine:2020} projects. Euclid proposed to use $KTaO_3$ as a tuning element, a ferroelectric crystal that exhibits excellent tuning parameters and very low loss tangent ($\tan \delta\simeq10^{-5}$ at X-band, while for the standard Strontium Titanate (SrTiO$_3$ or STO) material it is $\sim10^{-3}$) in the cryogenic temperature range $0.1$ - $10$~K \cite{Geyer:2005}. Euclid carried out electromagnetic simulations to study the KTO permittivity value and its variation inside an $8$~GHz test cavity \cite{EuclidTechlabs}. Euclid is currently working to manufacture a small prototype of the test cavity with a ferroelectric element, and to characterize the small test cavity in the $2$ - $10$~K range in order to experimentally verify the expected permittivity value, its tuning range, and its loss tangent \cite{EuclidTechlabs}.\\

Ferroelectrics are non-linear dielectrics with high relative permittivity values widely employed in high-density commercial decoupling capacitors, acoustic-electronic transducers, and MEMS \cite{Spartak}. Its permittivity can be modified by applying an external static electric field (or also varying the temperature), which is the most important attribute of some perovskites that make them attractive for agile microwave components (such as varactors, tunable RF filters, and phase shifters).\\

There are more than 300 ferroelectric materials available, but the most widely used is the Barium Strontium Titanate (Ba$_x$Sr$_{1-x}$TiO$_3$ or BST) which operates in paraelectric phase at room temperatures ($\sim297$~K) \cite{Ahmeda:2015,Kanareykina:2009}. For cryogenic temperatures, the STO was previously used by the axion research group of ADMX-Fermilab that has investigated ferroelectrics as a tuning element for the ADMX project \cite{Braine:2020}. The Fermilab's project \cite{Bowring:2018,Fermilab:2017} focused on exploiting the novel electronic properties of nonlinear dielectric materials such as Strontium Titanate to build electronically tunable detectors for axion dark matter experiments. The group has produced and tested a number of thick-film STO samples on quartz and sapphire substrates using several film deposition techniques. The frequency shifts and the corresponding dielectric permittivity changes were not of the expected magnitude \cite{Bowring:2018,Fermilab:2017}. A tuning range of less than $0.1\%$ was achieved by ADMX (very far from the tuning ranges achieved in many haloscope experiments using mechanical tuning systems, $>1\, \%$ \cite{Stern:2015,Boutan:2018,Zhong:2018,Choi:2021}), and the losses produced a reduction of more than $90\, \%$ in the unloaded quality factor. To the authors' knowledge, no other group has proposed an electronic tuning system without mechanical motions.\\

In this manuscript we propose novel electrical tuning systems for axion dark matter searches studied in the RADES project scenario. In contrast to mechanical tuning, electrical tuning can provide a system that is less prone to mechanical failures (it avoids movable parts in a cryogenic environment, for example) and thus complements and expands existing techniques. Another advantage of this kind of tuning for multi-cavity systems is the ability to independently adjust the different cavity frequencies and interresonator coupling values to maintain the correct modal structure and electromagnetic field pattern (which could improve also the form factor). In addition, the mechanical tuning does not behave well in scalability when these haloscopes are designed at higher frequencies (for example from X-band to Ku-band). This issue might be solved with the electrical tuning systems proposed in this work.\\

Different tuning technologies have been studied by the authors to implement an electrical tuning system in our haloscopes: dielectrics, ferromagnetics, ferroelectrics, liquid crystals, piezoelectrics, microelectromechanical systems (MEMS), composite ceramics based on mixtures of ferroelectrics and semiconductors (varactors) \cite{Cameron}. However, due to the high requirements imposed in an axion detection system (cryogenic temperatures, high static magnetic field level, and low losses), the most promising is the ferroelectric technology, which could work in this environment. This kind of dielectric material provides a permittivity change with temperature or bias DC voltage, so the idea is to load the haloscope with such material and then modify its relative electric permittivity $\varepsilon_r$ to produce a frequency shift. As a first order approximation, we can use the equation of the resonant frequencies of a completely filled homogeneous rectangular cavity:
\begin{equation}\label{eq:frmnl}
    f_{mnl} = \frac{c}{2 \, \sqrt{\varepsilon_r \, \mu_r}} \, \, \sqrt{\left(\frac{m}{a}\right)^2+\left(\frac{n}{b}\right)^2+\left(\frac{l}{d}\right)^2},
    \begin{alignedat}{2}
    \qquad m=0,1,2,...\\
    \qquad n=0,1,2,...\\
    \qquad l=1,2,3,...\\
    \qquad \text{$m$ and $n$ not zero simultaneously}
    \end{alignedat}
\end{equation}
where $f_{mnl}$ is the resonant frequency for a $TE_{mnl}$ mode in a rectangular cavity, $c$ the speed of light in vacuum, $\mu_r$ the relative magnetic permeability, and $a$, $b$, and $d$ the dimensions of the rectangular cavity (width, height and length). Equation~\ref{eq:frmnl} is also valid for $TM_{mnl}$ modes, but the indexes $m$ and $n$ cannot be zero, and only the index $l$ can be zero. The $TE_{101}$ fundamental resonance has been used in RADES for a dipole magnet scenario (CAST at CERN \cite{CAST:2020rlf}). We use this mode to conceive the haloscopes proposed in this work. As it can be seen, the resonant frequency is inversely proportional to the square root of the relative permittivity ($f_{mnl}\propto \frac{1}{\sqrt{\varepsilon_r}}$), so the higher the ferroelectric relative permittivity, the lower the haloscope frequency (and the axion mass to be scanned). This permittivity variation also modifies the distribution of the electromagnetic field inside the cavity, which is the effect that is exploited in this work for the tuning. In section~\ref{KTOasTuner} this effect is studied.\\

\begin{figure}[h]
\centering
\begin{subfigure}[b]{0.49\textwidth}
         \centering
         \includegraphics[width=1\textwidth]{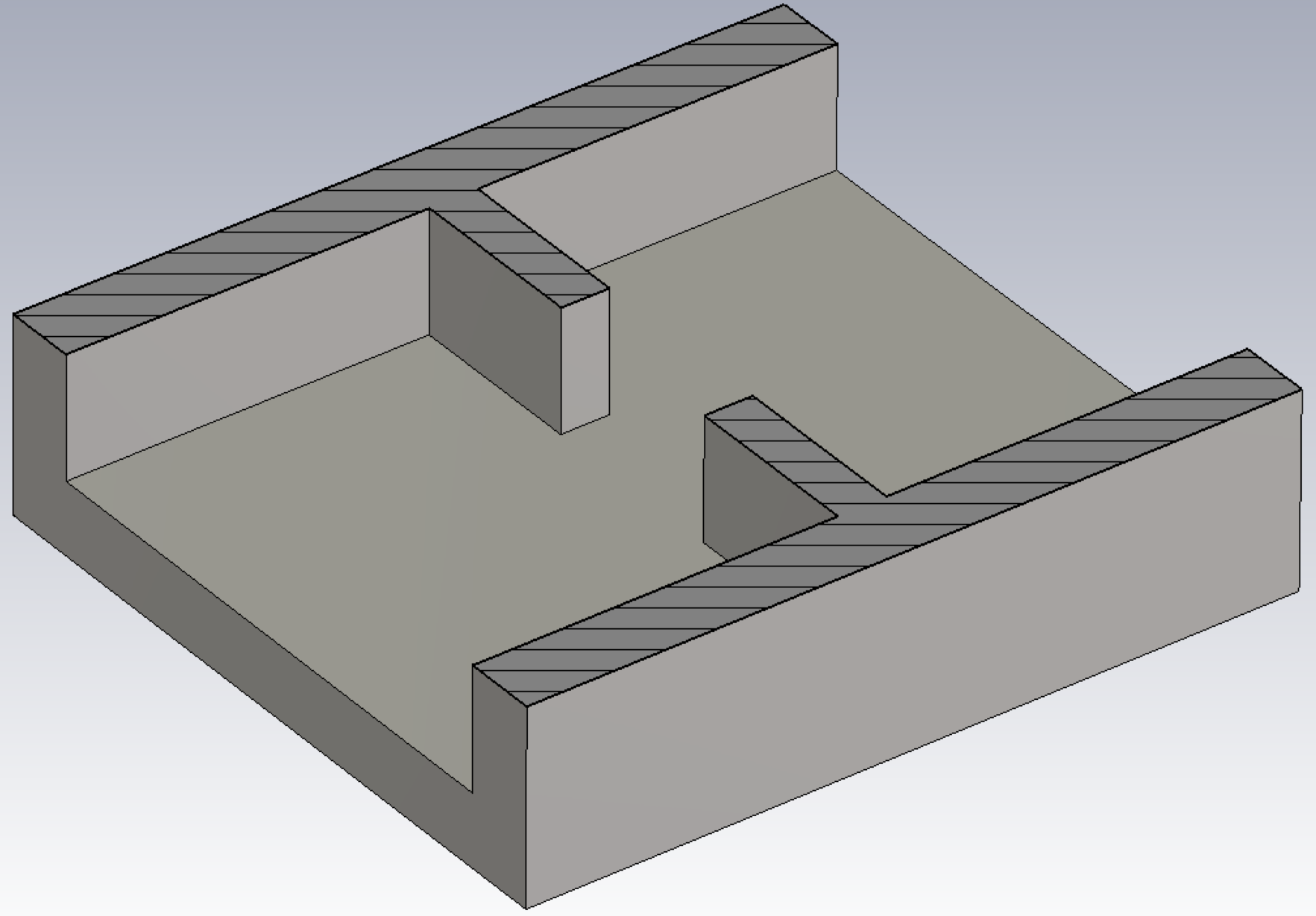}
         \caption{}
         \label{fig:ind}
\end{subfigure}
\hfill
\begin{subfigure}[b]{0.49\textwidth}
         \centering
         \includegraphics[width=1\textwidth]{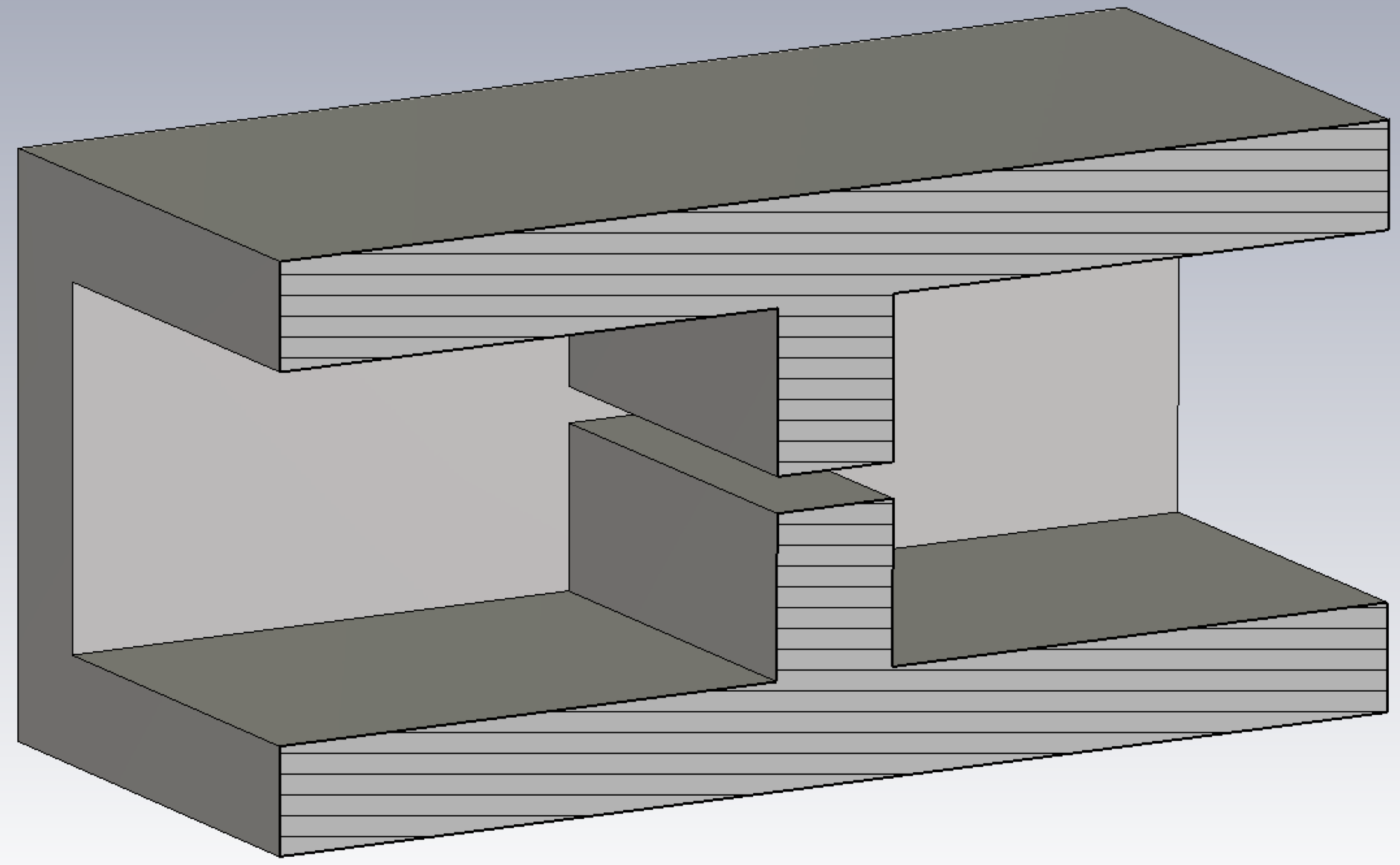}
         \caption{}
         \label{fig:cap}
\end{subfigure}
\caption{Two types of iris couplings: inductive window (a), and capacitive window (b). The pictures show the symmetric half of each one, being the dashed regions the symmetry planes.}
\label{fig:IrisCouplings}
\end{figure}

The second idea studied in this paper due to the high variation in the permittivity provided by ferroelectric films, is its use as interresonator coupling elements in multicavity haloscopes. In the RADES group \cite{RADESreviewUniverse} rectangular waveguide cavities connected by iris couplings have been used in the past for the haloscope designs. The coupling between cavities can be obtained by inductive or capacitive irises (see Figure~\ref{fig:IrisCouplings}), which also determines the sign of the coupling \cite{Cameron}.
Thus, the size of the irises controls the amount of coupling between adjacent cavities. This interresonator coupling is mainly used for improving the mode separation between the axion resonance and its neighbours, so a good coupling element should be able to effectively control the amount and the sign of the coupling between adjacent cavities.\\

These interresonator irises tend to give problems during manufacturing, especially if both types (inductive and capacitive) are combined in a single structure \cite{RADES_paper2}. In this case, the fabrication requires fully 3D machining techniques involving several cut planes. In previous RADES experiments the poor contact between metallic surfaces has led to a degradation in the flowing of the surface currents, and therefore to the deterioration of the unloaded quality factor. The use of ferroelectric films provides an efficient way to avoid these drawbacks, as both signs of the couplings can be obtained very easily with KTO films placed inside the fabricated rectangular housing without the need to cut at regions of high surface current levels. \\

On the other hand, thanks to the ability to adjust the permittivity, these new interresonator couplings allow fine tuning of the coupling due to possible errors caused by manufacturing tolerances and misalignment of the films inside the rectangular housing. This provides an efficient adjustment mechanism of the mode separation, which could otherwise cause an important deterioration of the response in a regular haloscope cavity.\\

In section~\ref{KTOasTuner} we first briefly review the main properties of ferroelectric materials, and then we describe our novel proposal to use them as tuning elements in cavity haloscopes for alleviating the issues reported in \cite{Bowring:2018,Fermilab:2017}, like the limited tuning range. In section~\ref{KTOasCoupler} the use of ferroelectric materials as interresonator coupling elements is explored for the design of a reconfigurable haloscope, that controls the mode separation. In section~\ref{Biasing} we comment some interesting practical considerations for the correct biasing of the ferroelectrics. Finally, in section~\ref{Conclusions} we expose our conclusions and discussion of future prospects.
\section{The KTO ferroelectric as a tuning element}
\label{KTOasTuner}
The KTO material is an incipient ferroelectric material with properties very similar to those of STO and Calcium Titanate (CaTiO$_3$ or CTO) \cite{Skoromets:2016}. Although KTO and STO have many similarities, the first  material maintains the cubic structure at very low temperatures just decreasing its loss tangent between room temperature and $5$~K although this is not the case for the STO material \cite{Skoromets:2016,Geyer:2005,Tagantsev:2003}. These lower losses make KTO single crystals interesting for the development of microwave tuning systems at cryogenic temperatures.\\

It should be noted that KTO single crystals have not been comprehensively studied in the $1$ - $4$~K and mK temperature ranges \cite{Geyer:2005}. In \cite{Geyer:2005}, the microwave dielectric properties of high purity KTO ferroelectrics have been measured at $T=4$~K using a cylindrical sample as $TE_{0n1}$ and quasi-$TE_{011}$ dielectric resonators. In comparison with single-crystal $SrTiO_3$, the $KTaO_3$ does not undergoes a phase transition at cryogenic temperatures. KTO remains paraelectric down to $5.4$~K, which is consistent with theoretical predictions. Thus, its dielectric loss tangent would be expected to generally decrease with decreasing temperature. This expected decrease of the dielectric loss tangent with temperature from $300$ to $4$~K over the frequency range from $1$ to $10$~GHz was demonstrated in \cite{Geyer:2005}. At zero bias, the dielectric constant increases to about $\varepsilon_r=4500$ on cryogenic cooling (from $\sim10$ to $4$~K), but saturates at liquid helium temperatures \cite{Geyer:2005}.\\

In summary, KTO at cryogenic temperatures could allow the development of an electronic tuning system with very low losses. Crystals of $99.99\, \%$ purity are commercially available in \cite{KTaO3}.
\subsection{Rectangular haloscope tuning with ferroelectrics}
\label{KTOatSides}
\begin{figure}[h]
\centering
\begin{subfigure}[b]{0.42\textwidth}
         \centering
         \includegraphics[width=1\textwidth]{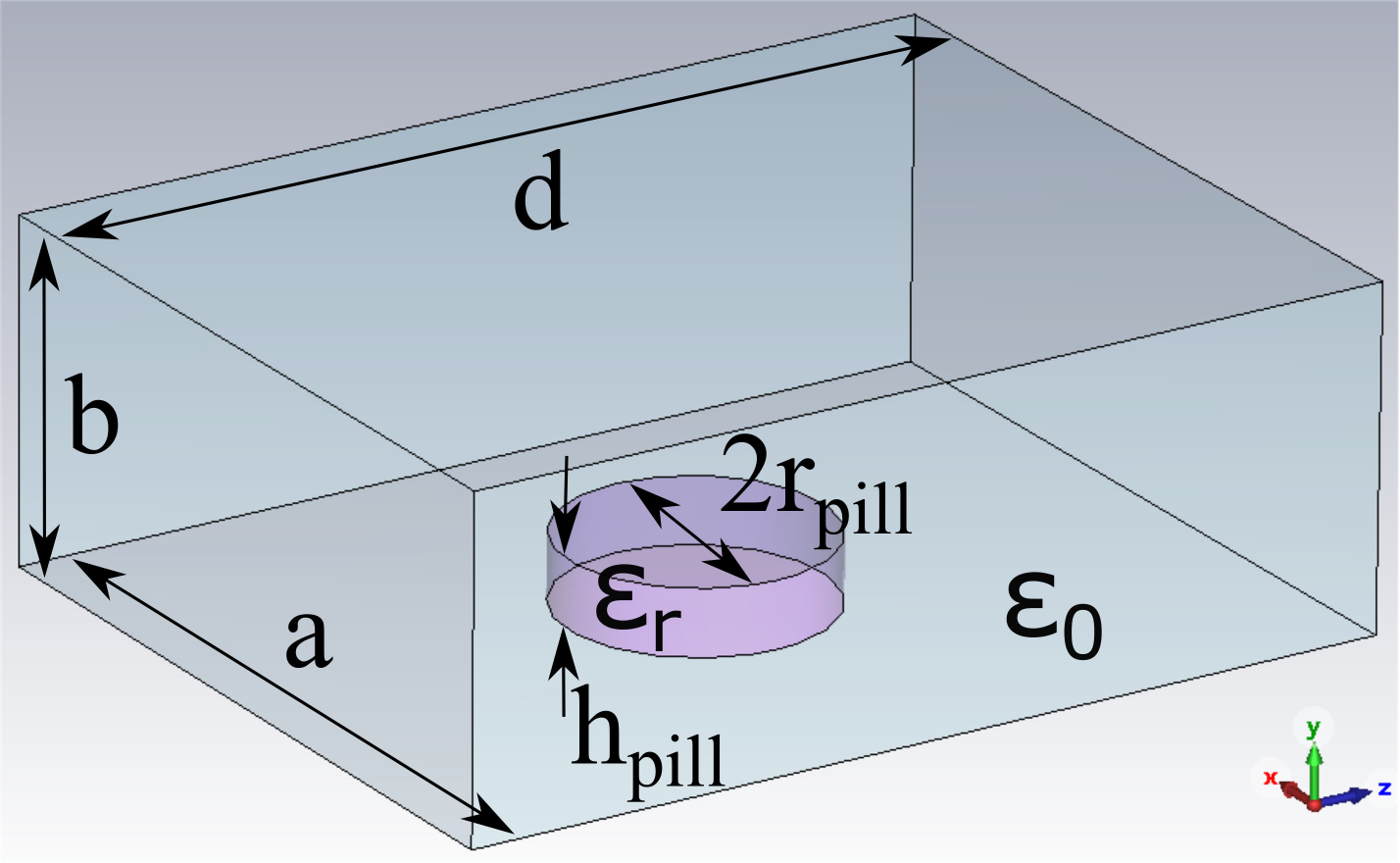}
         \caption{}
         \label{fig:1cav_KTOPillAtTheMiddle3Dview}
\end{subfigure}
\hfill
\begin{subfigure}[b]{0.56\textwidth}
         \centering
         \includegraphics[width=1\textwidth]{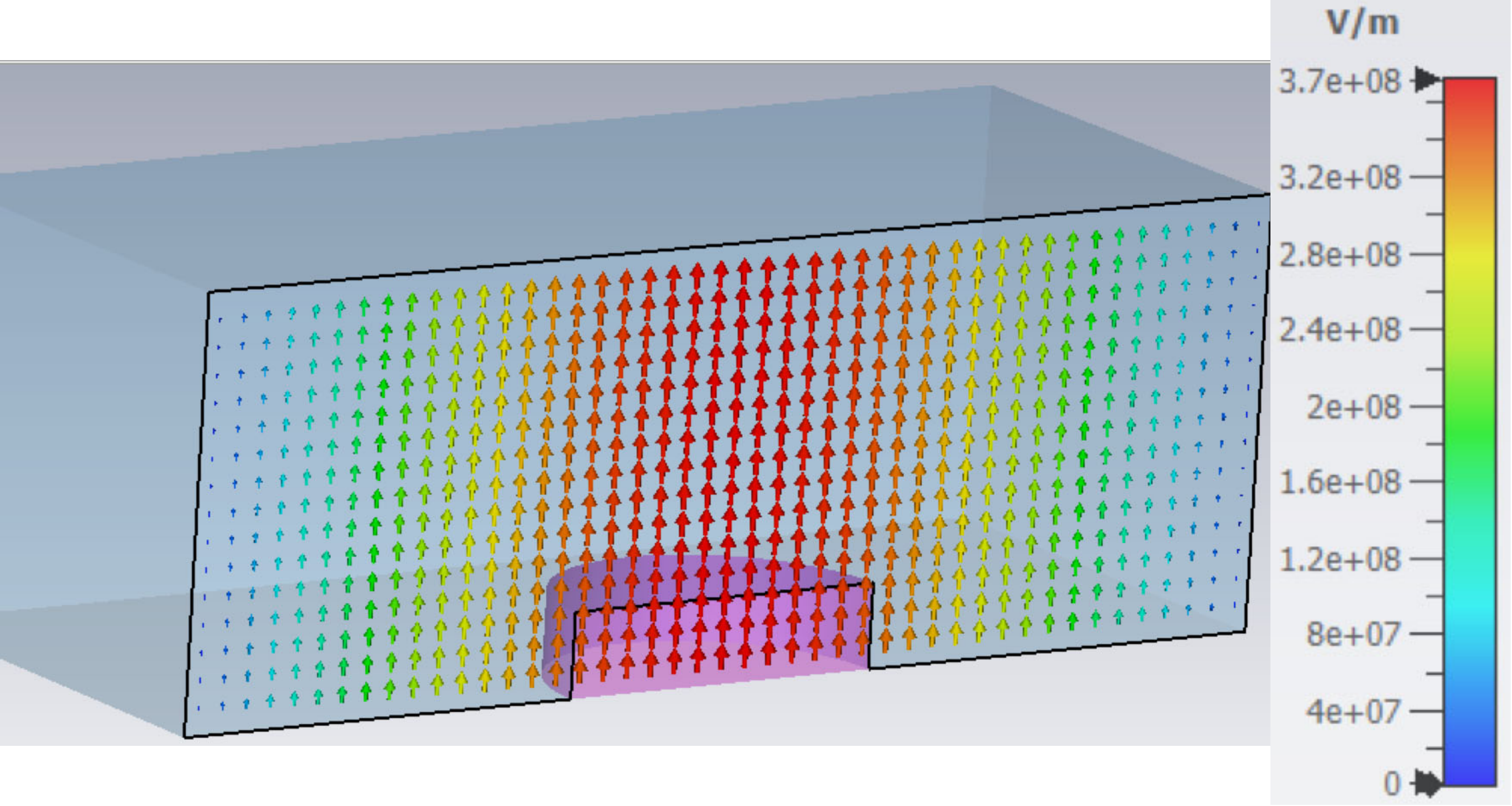}
         \caption{}
         \label{fig:1cav_KTOPillAtTheMiddleEfieldEps1}
\end{subfigure}
\hfill
\begin{subfigure}[b]{0.56\textwidth}
         \centering
         \includegraphics[width=1\textwidth]{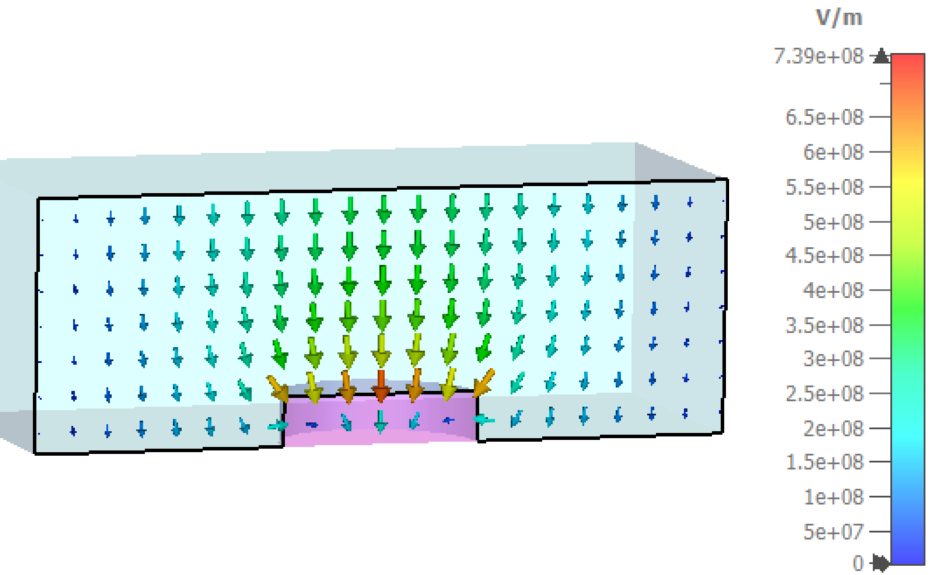}
         \caption{}
         \label{fig:1cav_KTOPillAtTheMiddleEfieldEps50}
\end{subfigure}
\caption{Cavity loaded with a dielectric pill (modeling a high permittivity material) at the centre: (a) 3D view, (b) electric field in a cross-section for $\varepsilon_r = 1$, and (c) for $\varepsilon_r = 50$. In this case, the dielectric behaves as a dielectric resonator. Simulation results obtained from the Computer Simulation Technology (CST) Studio Suite software \cite{CST}.}
\label{fig:1cav_KTOPillAtTheMiddle}
\end{figure}

Once we have clarified the properties of these materials, we describe the proposed system to use KTO as tuning elements in rectangular cavity haloscopes. For this purpose, the first step is to find the best position of KTO objects (with a particular geometry) inside the cavity, in order to provide a good frequency tuning range and to reduce the impact on other parameters such as the form factor and the quality factor. The optimal position of such ferroelectric items must be sought in order to avoid a strong concentration of electromagnetic field within and around the KTO. This task represents a great challenge due to the very high dielectric constant value of KTO, specially at low temperatures. For example, if we place a high-permittivity pill at the bottom centre of a rectangular cavity (see Figure~\ref{fig:1cav_KTOPillAtTheMiddle3Dview}) the electric field is curved and tends to be concentrated in the vicinity of the material for high values of the dielectric constant (in this case 50), as shown in Figures~\ref{fig:1cav_KTOPillAtTheMiddleEfieldEps1} and \ref{fig:1cav_KTOPillAtTheMiddleEfieldEps50}; this behaviour is denoted as dielectric resonator which has been extensively studied in the microwave engineering literature \cite{Kajfez:1986}. In Table~\ref{tab:1cav_KTOPillAtTheMiddle} the resonant frequency, the unloaded quality factor $Q_0$ and the form factor $C$ are listed for several dielectric constant values in this system. Note how the form factor for $\varepsilon_r = 1$ is equal to the one for the cavity without dielectrics for the $TE_{101}$ mode: $C=64/\pi^4=0.657$, which can be obtained analytically from equation~\ref{eq:C}. Despite the good frequency shift achieved for $\varepsilon_r = 50$ it can be observed how the quality and form factors are significantly decreased. This result clearly indicates that this position and shape for the KTO, which in practise has very large dielectric constant values (close to $\varepsilon_r=4500$), should be avoided.\\

\begin{figure}[h]
\centering
\includegraphics[width=0.8\textwidth]{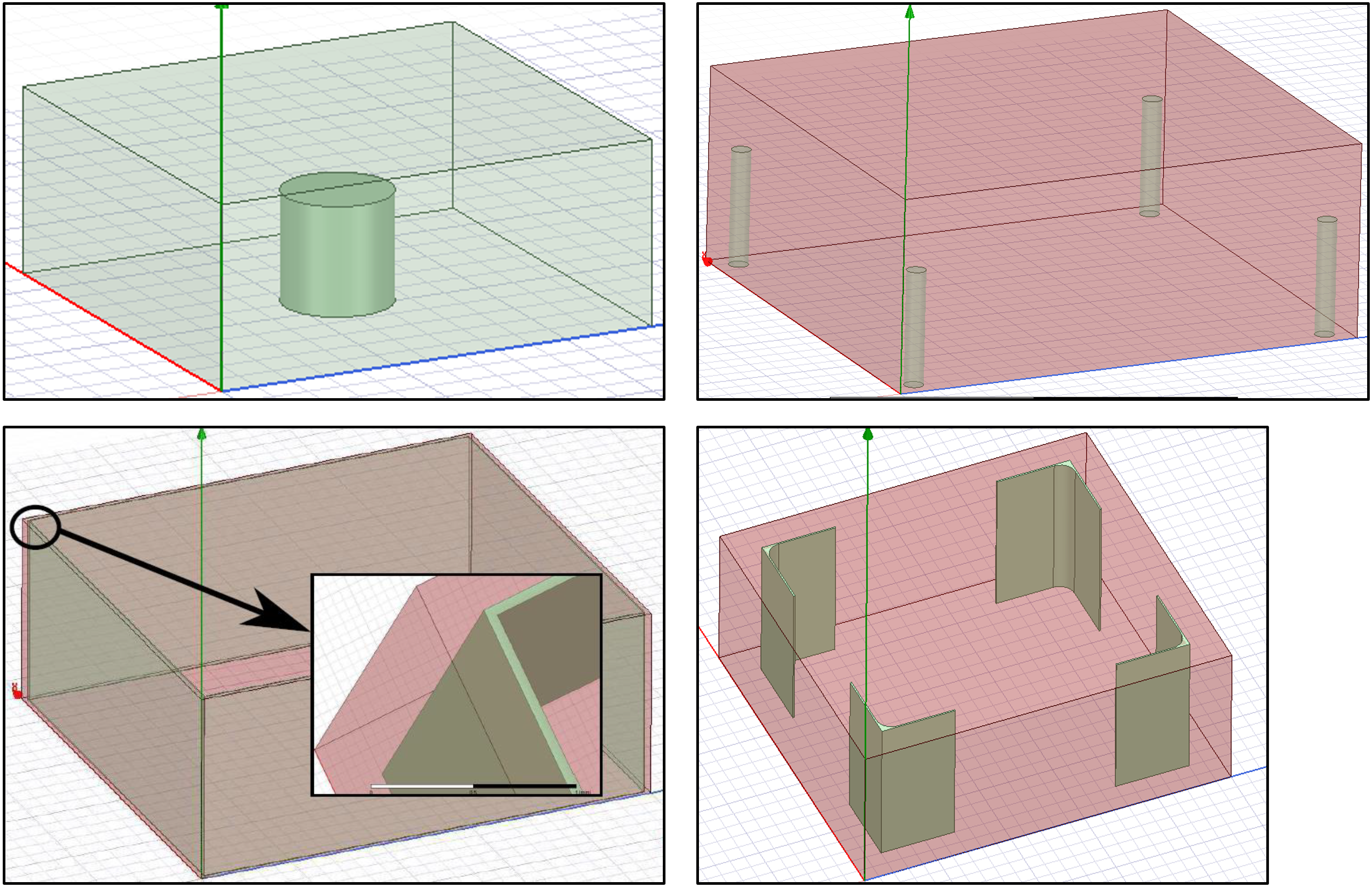}
\caption{\label{fig:FerroelectricShapes_and_PositionsStudied} 
Examples of geometrical shapes and positions studied for the KTO in our tuning system. These configurations provide non-optimum results in terms of quality factor, form factor and tuning frequency range, and were discarded as viable solutions for tunable haloscopes.}
\end{figure}

\begin{figure}[h]
\centering
\begin{subfigure}[b]{0.42\textwidth}
         \centering
         \includegraphics[width=1\textwidth]{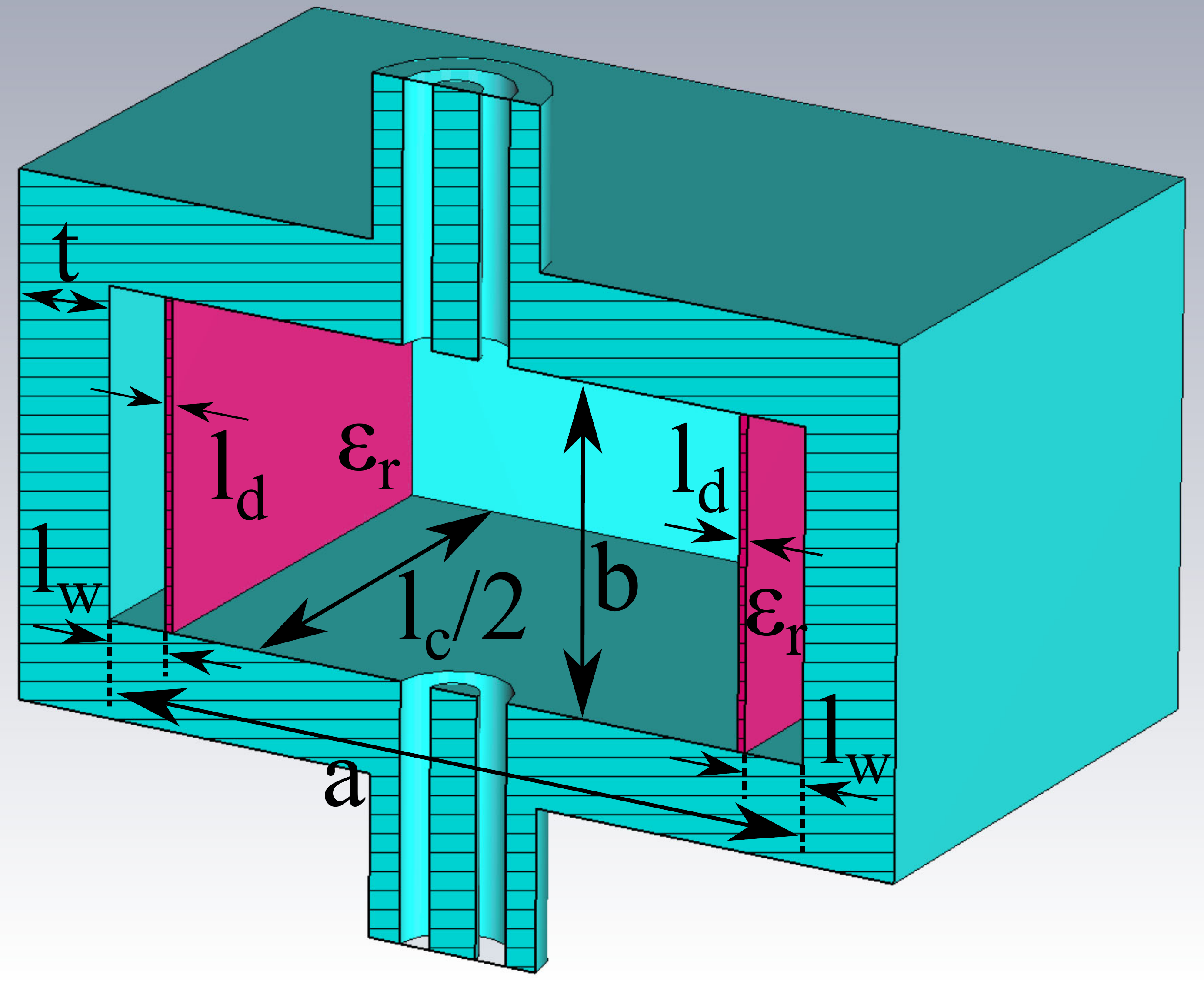}
         \caption{}
         \label{fig:1cav_KTOatSides_Model}
\end{subfigure}
\hfill
\begin{subfigure}[b]{0.49\textwidth}
         \centering
         \includegraphics[width=1\textwidth]{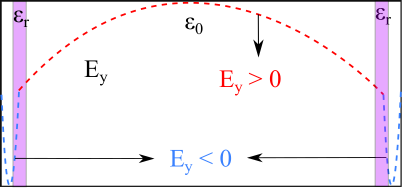}
         \caption{}
         \label{fig:1cav_KTOatSides_Efield_Transverse}
\end{subfigure}
\caption{(a) Model based on a resonant cavity with two coaxial ports. The picture shows the symmetric half, being the dashed region the symmetry plane. Ferroelectric KTOs are depicted in pink colour. The length of the cavity $l_c$ is adjusted to operate at the central working frequency. $l_w$ is the distance of the ferroelectric film to the wall and $t$ is the thickness of the cavity housing. The input/output coupling is controlled by the coaxial pin connector length (which is based on SMA type). (b) Transverse image of the vertical component of the electric field of the {\em modified}-$TE_{301}$ mode at the resonant frequency cavity with the outermost lobes (shown in blue dashed lines) inside the KTOs; note that the electric field is positive in the central region (shown in red dashed lines) whereas it becomes negative inside the dielectric media and beyond them (see blue dashed lines). Note that due to the strong concentration of the lateral lobes inside the KTOs, the {\em modified}-$TE_{301}$ strongly resembles the original $TE_{101}$ mode in a rectangular cavity.}
\label{fig:1cav_KTOatSides}
\end{figure}

\begin{table}[h]
\begin{tabular}{|c|c|c|c|}
\hline
$\varepsilon_r$ & $f_r$ (GHz) & $Q_0$ & $C$ \\ \hline\hline
1 & 8.406 & 45319 & 0.657 \\ \hline % Actually C = ~0.654
2 & 8.249 & 45223 & 0.624 \\ \hline
5 & 8.070 & 44923 & 0.589 \\ \hline
10 & 7.972 & 44569 & 0.573 \\ \hline
20 & 7.903 & 44004 & 0.554 \\ \hline
30 & 7.865 & 43005 & 0.540 \\ \hline
40 & 7.819 & 40110 & 0.509 \\ \hline
50 & 7.710 & 30135 & 0.386 \\ \hline %% eps_R = 60 --> fr = 7.38 GHZ
\end{tabular}                       %% eps_R = 70 --> fr = 6.919 GHZ, C = 0.031
\centering
\caption{\label{tab:1cav_KTOPillAtTheMiddle} Characteristics of a rectangular cavity (with $a=22.86$~mm, $b=10.16$~mm and $d=28.5$~mm) loaded with a lossless dielectric pill ($r_{pill} = 4$~mm,  $h_{pill} = 2$~mm) placed at the middle of the cavity (modeling a high permittivity material) for several dielectric constant values. An electric conductivity $\sigma=2\times 10^9$ S/m is assumed for the metallic housing. Simulation results obtained from CST \cite{CST}.}
\end{table}
After performing manually extensive optimization operations involving geometrical shapes and positions within the cavity (see examples in Figure~\ref{fig:FerroelectricShapes_and_PositionsStudied}), we finally obtained a tuning system concept based on thin ($l_d\simeq500$ microns, being $l_d$ the thickness of the KTO) ferroelectric films, located close to the cavity side walls (see Figure~\ref{fig:1cav_KTOatSides_Model}\footnote{Note here how the thickness of the metallic housing walls has been displayed, while for the rest of the models in the previous figures it is omitted.}) that provides good results in terms of quality factor, form factor and tuning frequency range.\\

The haloscope depicted in Figure~\ref{fig:1cav_KTOatSides_Model} is a single cavity with two KTO films separated $1.8$~mm from the side walls. For this example, the WR-90 rectangular waveguide section ($a= 22.86$~mm and $b=10.16$~mm), which works at X-band frequencies (our frequency region for this study), is used.\\

The analytical study of this cavity is not straightforward since a numerical resolution of a  transverse resonance equation is needed to model a structure containing a partially loaded rectangular waveguide section~\cite{pozar}. As an alternative, we perform the design using the electromagnetic (EM) simulation tool CST \cite{CST} on the frequency domain. Another issue that is found in the design of the structure is the high computational cost needed for the numerical characterization due to the presence of objects with a very high dielectric constant which leads to a significant number of mesh tetrahedrons cells in the dielectric material in order to achieve good numerical accuracy. The solution for this drawback was the use of adaptive meshing in CST, and the consideration of electrically very thin ferroelectric films. Alternatively, the structure was also modelled using FEST3D \cite{FEST3D}, a software tool which employs an integral equation technique efficiently solved by the Method of Moments (MoM) and the Boundary Integral-Resonant Mode Expansion (BI-RME) method. The use of these two full-wave modal electromagnetic simulators serves as validation of the numerical results obtained in the different designs of the ferroelectric haloscopes.\\

The key idea to conceive this kind of system is based on starting with the $TE_{301}$ cavity mode, which is perturbed by the high KTO permittivity value to make it to resemble the original $TE_{101}$ mode, increasing the form factor. The concept is shown in Figure~\ref{fig:1cav_KTOatSides_Efield_Transverse}. This new mode is called the $modified-TE_{301}$ mode, where two of the three variations of the electric field along the width axis are very narrow due to the high permittivity value of the KTO films (see blue dashed lines in Figure~\ref{fig:1cav_KTOatSides_Efield_Transverse}), while the central lobe is stretched along most of the vacuum section of the waveguide (this is depicted with red dashed lines in Figure~\ref{fig:1cav_KTOatSides_Efield_Transverse}). To achieve this, the electrical thickness of the ferreoelectric films must be close to the guided half-wavelength since this is the distance between two minimums of the transverse standing wave pattern. This is calculated as \cite{pozar}:
\begin{equation}
\label{eq:ld}
l_d=\frac{\lambda_{g}^{KTO}}{2}=\frac{\lambda_{0}}{2 \, \sqrt{\varepsilon_r} \, \sqrt{1-(f_{c}^{KTO} \, / \, f_0)^2}}=\frac{c}{2 \, f_0 \, \sqrt{\varepsilon_r} \, \sqrt{1-(f_{c}^{KTO} \, / \, f_0)^2}},
\end{equation}
where $\lambda_{g}^{KTO}$ is the guided wavelength at the KTO region, $\lambda_{0} = c/f_0$ the free-space wavelength, $f_{c}^{KTO} = c/(2l_c\sqrt{\varepsilon_r})$ the cut-off frequency of the $TE_{10}$ mode at the KTO region\footnote{The expression of $f_{c}$ has generally the $a$ (cavity width) parameter in the denominator. However, in this case we use the cavity lenght ($l_c$) because a transverse resonance equation analysis is being employed for considering the $x$-axis as the propagation axis for the $TE_{10}$ mode, and thus model the standing wave pattern along this axis.}, and $f_0$ the operation frequency.\\

The variation of the KTO permittivity value effectively controls the position of the lateral lobes of the {\em modified}-$TE_{301}$, providing a good tuning range as will be proved in the next subsection. At the same time, it avoids the reduction of the form factor due to the small region where the electric field takes negative values (see blue dashed line in Figure~\ref{fig:1cav_KTOatSides_Efield_Transverse}).\\

The relative permittivity range assumed in this work varies from $\varepsilon_r = 3000$ to $5000$. However, this concept is scalable to any other situation which may be useful in the case that manufacturing tolerances of the ferroelectric material result in a different useful range. Also, in section~\ref{Introduction} we commented that the KTO losses are $\tan \delta = 10^{-5}$; however $\tan \delta = 10^{-4}$ is used in simulations to be more conservative in order to take into account future extra losses in real prototypes. For the electric conductivity of the metallic housing copper, a cryogenic value of $\sigma=2\times 10^9$ S/m is employed. It provides a $Q_0$ value around $4\times10^5$ for the rectangular cavity used in this work \cite{RADES_paper2}.\\

\subsection{Results for the KTO placed at the cavity side walls}
\label{KTOatSidesResults}
As an example, we have optimized the model from Figure~\ref{fig:1cav_KTOatSides_Model} at the operation frequency $f_0=8.5$~GHz. This leads, using equation~\ref{eq:ld}, to a KTO thickness from $l_d=250 \, \mu$m (for $\varepsilon_r=5000$) to $l_d=322 \, \mu$m (for $\varepsilon_r=3000$), as initial values. On the other hand, the length of the cavity $l_c$ is set to have a resonant frequency of $8.5$~GHz for the $TE_{101}$ mode, without ferroelectrics, that is, $l_c=26.97$~mm with a cavity width of $a=22.86$~mm and height $b=10.16$~mm.\\

\begin{figure}[h]
\centering
\begin{subfigure}[b]{0.49\textwidth}
         \centering
         \includegraphics[width=1\textwidth]{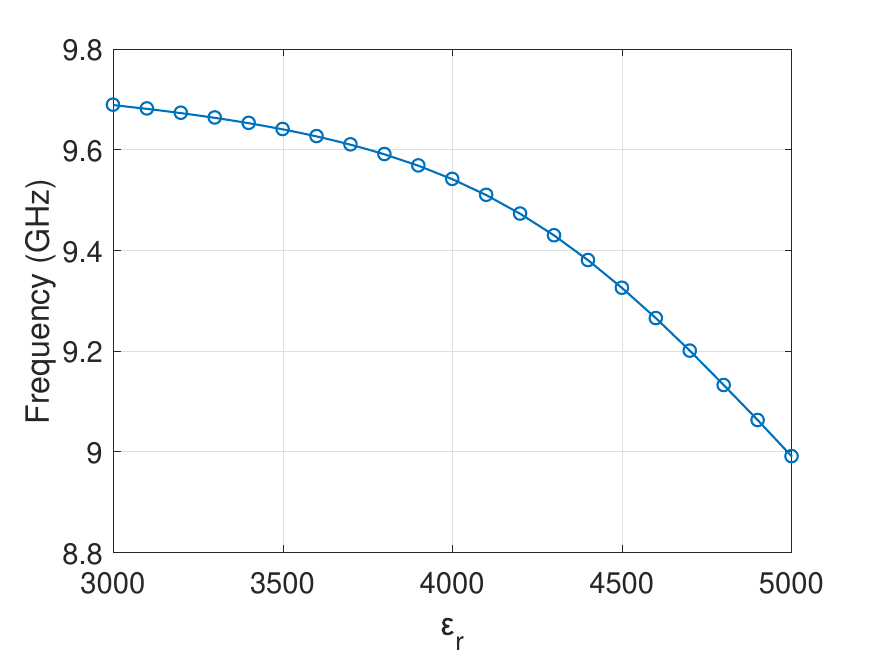}
         \caption{}
         \label{fig:1cav_KTOatSides_TuningRange}
\end{subfigure}
\hfill
\begin{subfigure}[b]{0.49\textwidth}
         \centering
         \includegraphics[width=1\textwidth]{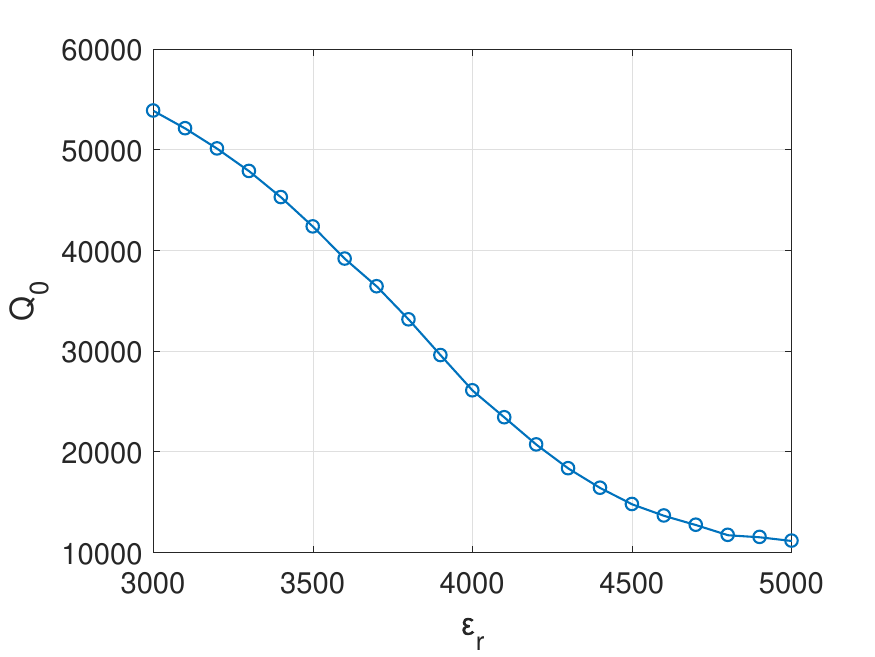}
         \caption{}
         \label{fig:1cav_KTOatSides_Q0Range}
\end{subfigure}
\hfill
\begin{subfigure}[b]{0.49\textwidth}
         \centering
         \includegraphics[width=1\textwidth]{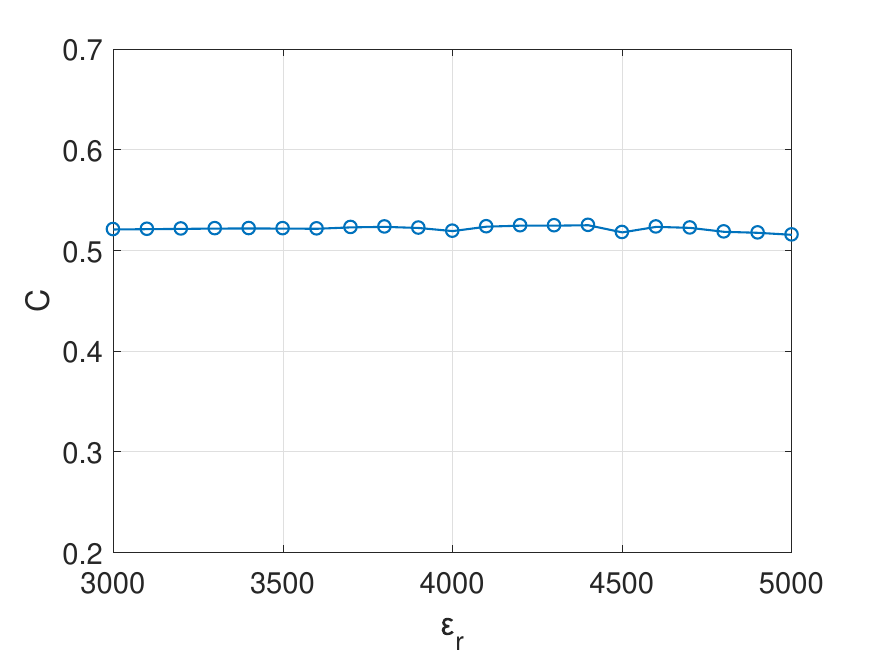}
         \caption{}
         \label{fig:1cav_KTOatSides_CRange}
\end{subfigure}
\hfill
\begin{subfigure}[b]{0.49\textwidth}
         \centering
         \includegraphics[width=1\textwidth]{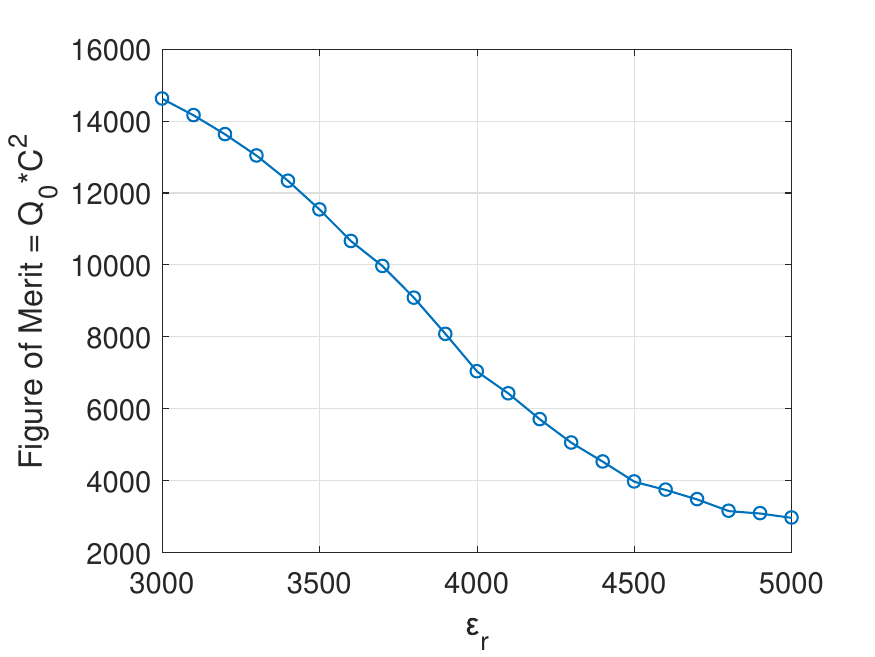}
         \caption{}
         \label{fig:1cav_KTOatSides_FOMRange}
\end{subfigure}
\caption{Parameters of the model from Figure~\ref{fig:1cav_KTOatSides} as a function of the dielectric constant for $\tan \delta = 10^{-4}$. (a) Resonant frequency. (b)
Unloaded quality factor. (c) Form factor. (d) Figure of Merit. Simulation results obtained from CST \cite{CST}.}
\label{fig:1cav_KTOatSides_TuningRanges}
\end{figure}

Now, in order to select the best film thickness ($l_d$) and separation from the walls ($l_w$), an optimization process is followed, modifying slightly the values of $l_d$ and $l_w$ (see Figure~\ref{fig:1cav_KTOatSides}), to find the highest tuning range with the best quality and form factors. As it is explained in section~\ref{Introduction}, the parameters that are analysed for the performance of a haloscope design are $\kappa$, $V$, $C$ and $Q_0$. More specifically, following equation~\ref{eq:dmadt} and taking into account that $\kappa$ and $V$ remain unchanged during the design, a convenient figure of merit is $FM=Q_0\, C^2$. After performing an optimization process based on maximizing FM (as in \cite{RADES_paper1}) and the tuning range, an optimum design is obtained with $l_d = 235 \, \mu$m and $l_w = 1.88$~mm. In Figure~\ref{fig:1cav_KTOatSides_TuningRanges} the tuning range, unloaded quality factor, form factor and figure of merit parameters are shown for the previous configuration. With these values a $700$~MHz frequency ($7.2$~$\%$) tuning range is achieved with an almost constant $C=0.52$ for all the tuning range and with $Q_0\in[10^5,5.5\times10^5]$ (see Figure~\ref{fig:1cav_KTOatSides_TuningRanges}).\\

\begin{figure}[h]
\centering
\includegraphics[width=0.8\textwidth]{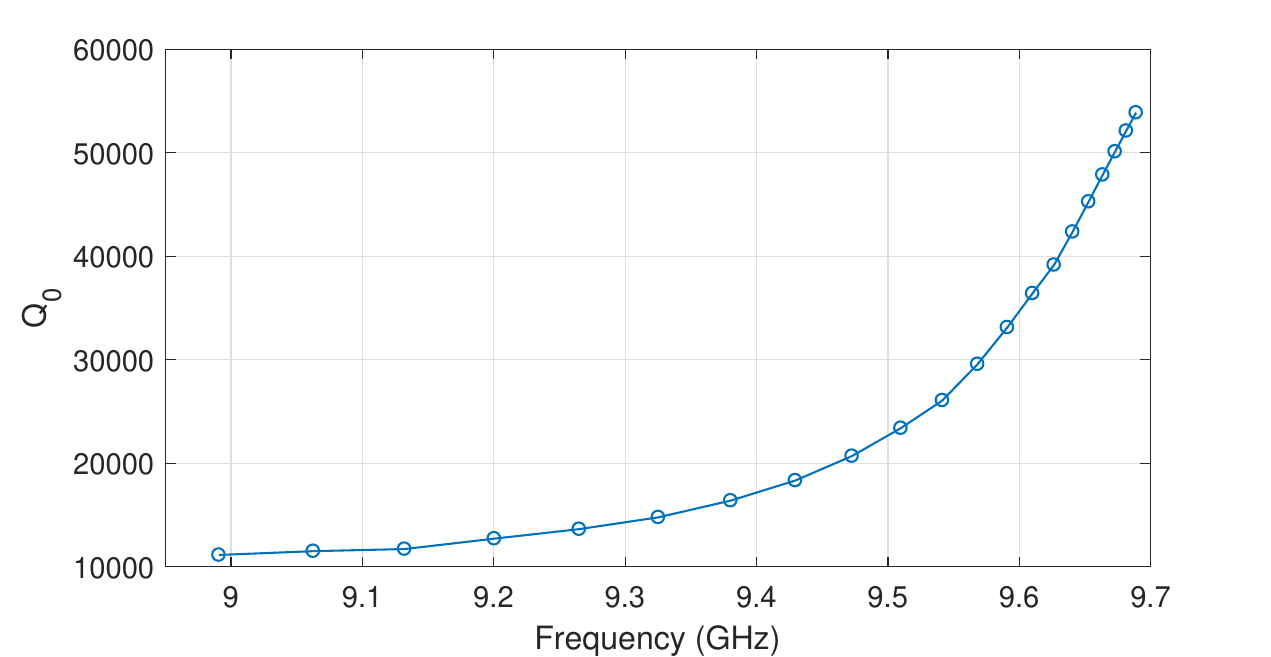}
\caption{\label{fig:1cav_KTOatSides_Q0vsFreq} 
Dependence of the unloaded quality factor $Q_0$ with the frequency in the model of Figure~\ref{fig:1cav_KTOatSides}. Simulation results obtained from CST \cite{CST}.}
\end{figure}

From the results displayed in Figure~\ref{fig:1cav_KTOatSides_TuningRange} and \ref{fig:1cav_KTOatSides_Q0Range}, it can be extracted the plot shown in Figure~\ref{fig:1cav_KTOatSides_Q0vsFreq} that depicts the dependence of the unloaded quality factor with the frequency: the higher the frequency shift (towards lower frequencies), the lower the quality factor.\\

If the drop in the unloaded quality factor is not acceptable, the system can be designed with a reduced range $\varepsilon_r \in [3000 , 4200]$, which provides a tuning range of $2.23\, \%$ ($216$~MHz) with an unloaded quality factor higher than $Q_0=20700$, which is a competitive value for axion searches at this frequency \cite{RADES_paper2}.\\

\subsection{Results for a multicavity haloscope}
\label{KTOatSidesResultsMulticavity}
\begin{figure}[h]
\centering
\begin{subfigure}[b]{0.59\textwidth}
         \centering
         \includegraphics[width=1\textwidth]{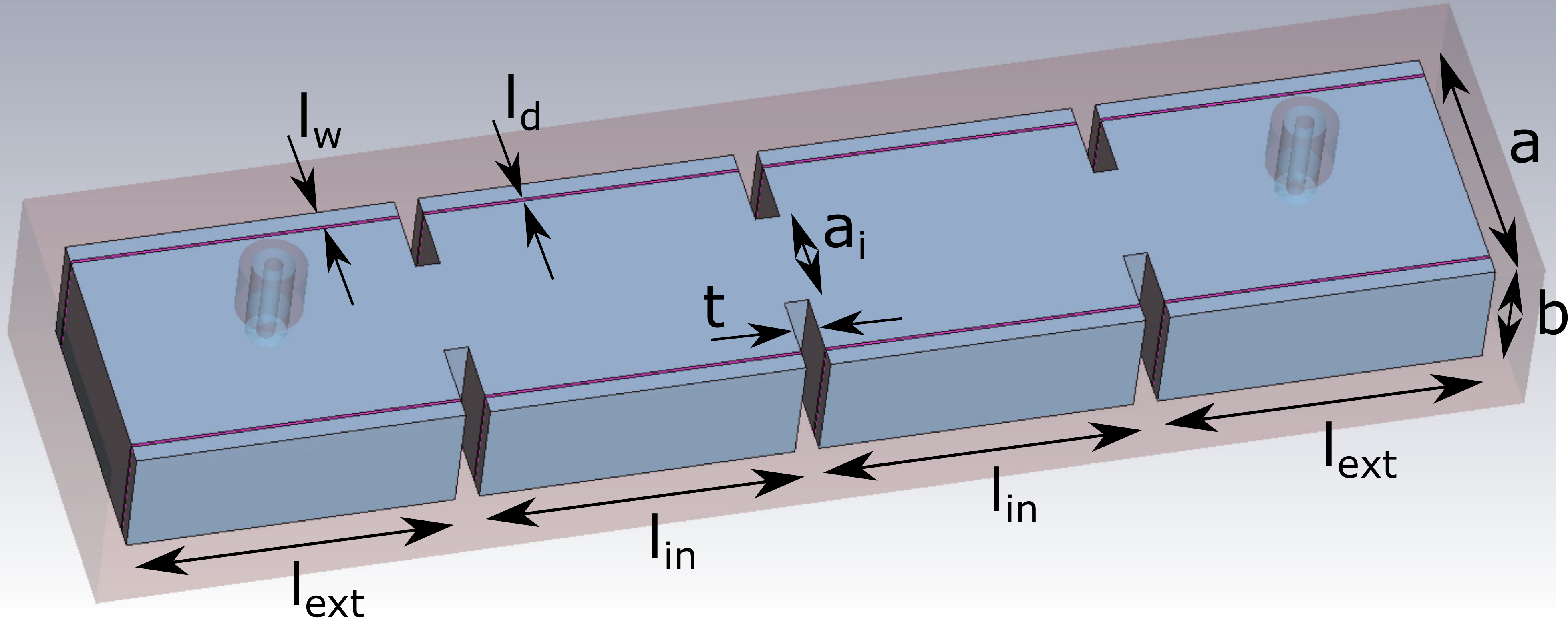}
         \caption{}
         \label{fig:4cav_KTOatSides}
\end{subfigure}
\hfill
\begin{subfigure}[b]{0.4\textwidth}
         \centering
         \includegraphics[width=1\textwidth]{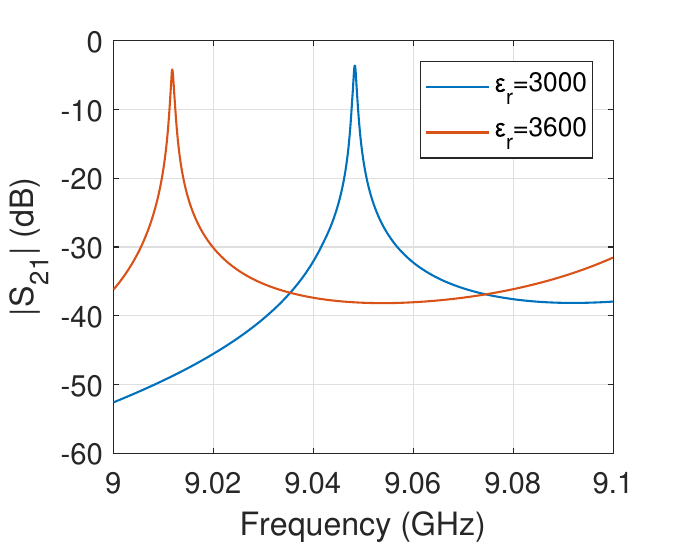}
         \caption{}
         \label{fig:4cav_KTOatSides_Sp}
\end{subfigure}
\caption{(a) Model based on an all-inductive four-subcavities haloscope with a coaxial port at each end cavity. (b) $S_{21}$ scattering parameter magnitude as a function of the frequency for two different values of the relative permittivity, showing a shift in the resonant frequency of $36.5$~MHz. The width and height of all the cavities are $a=22.86$~mm and $b=10.16$~mm, respectively, the length of the two end cavities is $l_{ext}=26.97$~mm, the length of the two inner cavities is $l_{in}=26$~mm, and the thickness and width of the three inductive irises are $t=2$~mm and $a_i=9$~mm, respectively. Simulation results obtained from CST \cite{CST}.}
\label{fig:4cav_KTOatSides_Model&Sp}
\end{figure}

A haloscope composed of four cavities coupled with inductive irises \cite{RADES_paper1} has been designed with this ferroelectric concept using the standard WR-90 rectangular waveguide section (see Figure~\ref{fig:4cav_KTOatSides}) based on $a=22.86$~mm and $b=10.16$~mm. In this case, we have employed in all subcavities a KTO thickness of $l_d = 250\, \mu$m with $l_w=1.5$~mm, obtained again from an optimization procedure based on maximizing FM and the tuning frequency range, as in the cavity from the previous section. Figure~\ref{fig:4cav_KTOatSides_Sp} shows the responses obtained for two different values of the relative permittivity, $3000$ and $3600$. These results show that the resonant frequency has shifted by $37$~MHz ($0.4\, \%$) which establishes the basis towards the demonstration that this system can be used for the design of real tunable haloscopes based on multicavity structures.\\

For this relative permittivity range $\varepsilon_r \in [3000,3600]$ we have obtained an unloaded quality factor change of $Q_0 \in [46066 , 38369]$, and a form factor variation of $C \in [0.509 , 0.49]$. It is worth noting that this permittivity range provides, in the first example (model of only one cavity shown in Figure~\ref{fig:1cav_KTOatSides_Model}), a figure of merit FM of the same order of magnitude. In addition, when $\varepsilon_r$ increases, it hardly affects to the form factor due to the small volume of the KTO.\\

\begin{figure}[h]
\centering
\begin{subfigure}[b]{0.29\textwidth}
         \centering
         \includegraphics[width=1\textwidth]{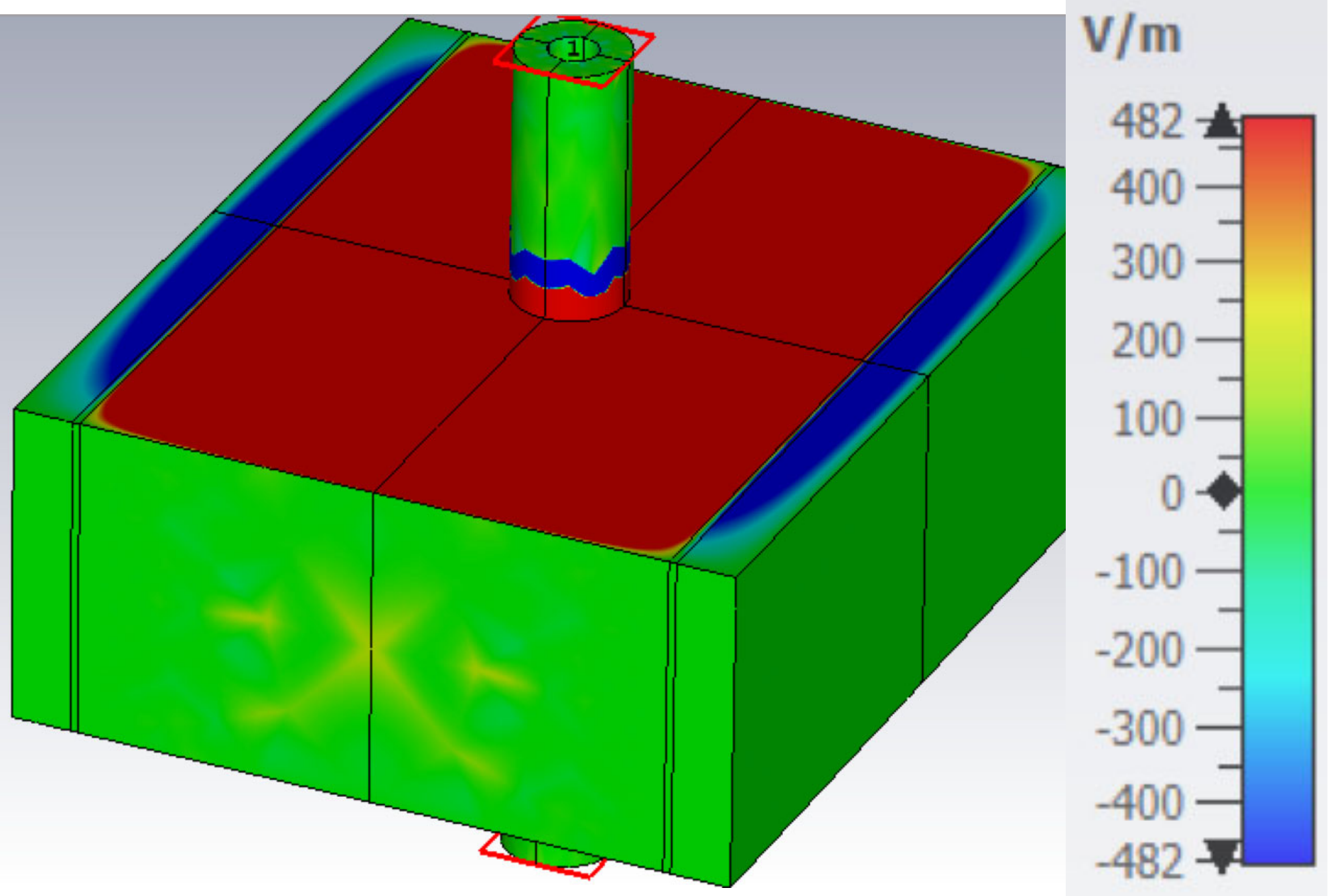}
         \caption{}
         \label{fig:1cav_KTOatSides_Efield}
\end{subfigure}
\hfill
\begin{subfigure}[b]{0.69\textwidth}
         \centering
         \includegraphics[width=1\textwidth]{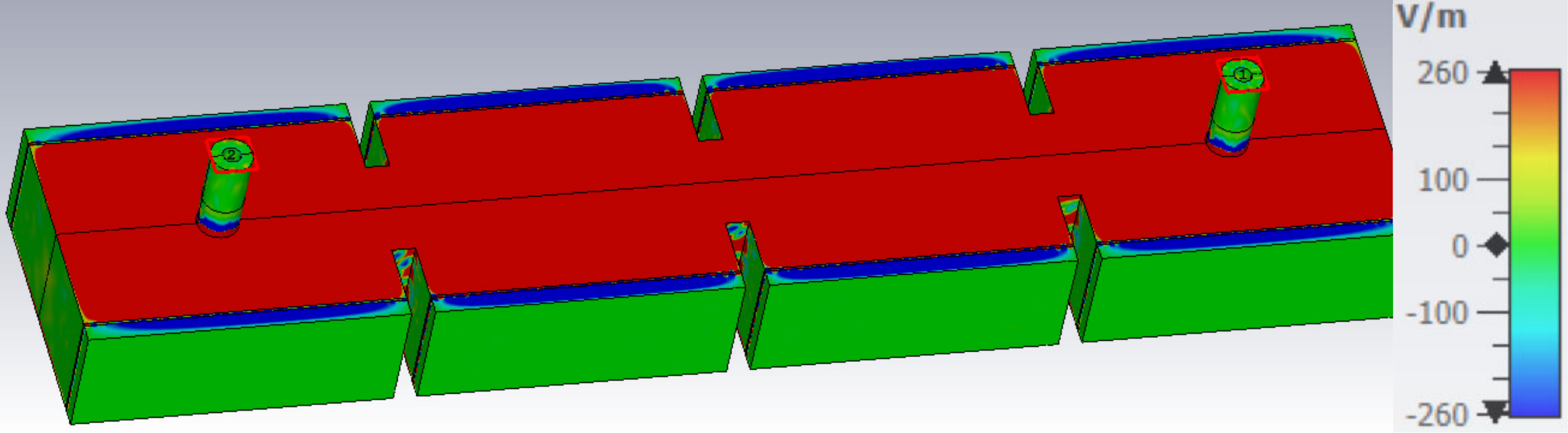}
         \caption{}
         \label{fig:4cav_KTOatSides_Efield}
\end{subfigure}
\hfill
\begin{subfigure}[b]{0.49\textwidth}
         \centering
         \includegraphics[width=0.59\textwidth]{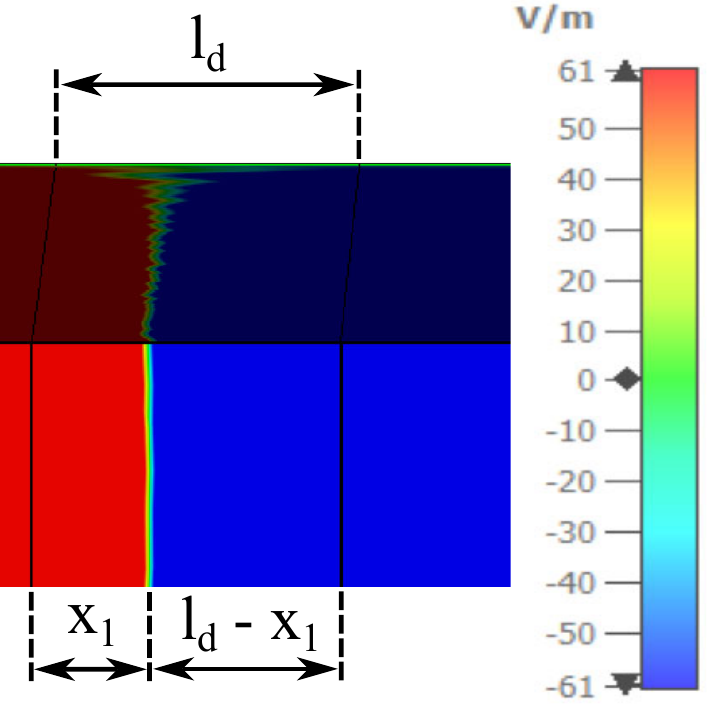}
         \caption{}
         \label{fig:1cav_KTOatSides_Efield_Zoom_er3000}
\end{subfigure}
\hfill
\begin{subfigure}[b]{0.49\textwidth}
         \centering
         \includegraphics[width=0.59\textwidth]{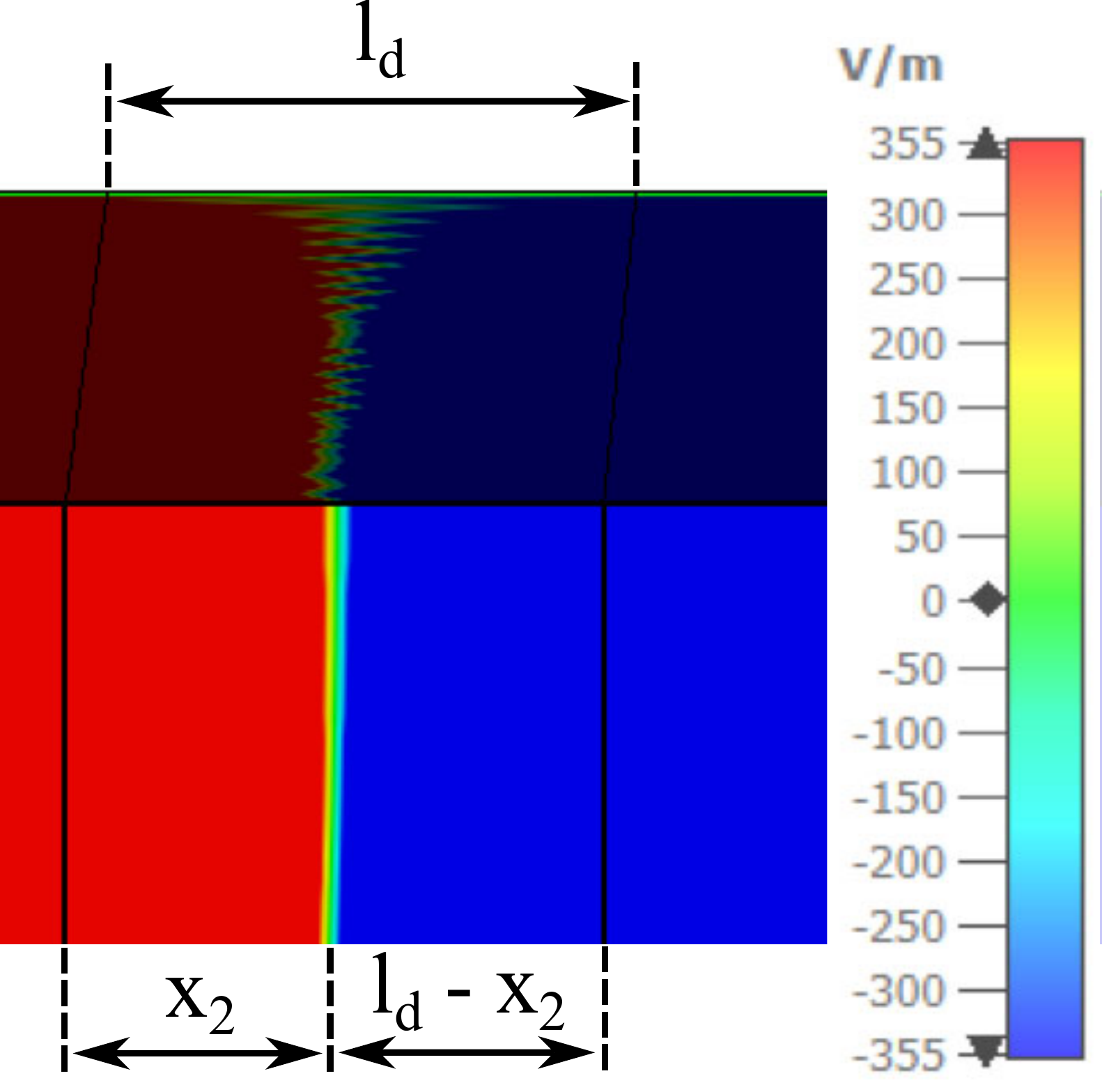}
         \caption{}
         \label{fig:1cav_KTOatSides_Efield_Zoom_er5000}
\end{subfigure}
\caption{Electric field (vertical component) for $\varepsilon_r=3000$ in the models (a) one-cavity and (b) four-subcavities. Zoom at the KTO area to observe the positive/negative transition of the {\em modified}-$TE_{301}$ mode inside the one-cavity model with (c) $\varepsilon_r=3000$ and (d) $\varepsilon_r=5000$. $x_1<x_2$, where $x_1$ and $x_2$ are the lengths of the positive E-field regions inside the KTO for $\varepsilon_r=3000$ and $\varepsilon_r=5000$, respectively.  Simulation results obtained from CST \cite{CST}.}
\label{fig:1&4cav_KTOatSides_Efield}
\end{figure}

Finally, in Figure~\ref{fig:1&4cav_KTOatSides_Efield} the electric field pattern for $\varepsilon_r=3000$ in both designs is observed: one cavity and four-subcavities models \footnote{In these images a strong colour scaling has been applied to correctly appreciate the negative electric field at the ferroelectric area. Without this scaling, small levels of electric field would be observed around the inductive couplings, indicating that this is the correct mode configuration.}. As it can be seen, the axion mode is the {\em modified}-$TE_{301}$ one for both cases. For higher permittivity values, the central lobe of the electric field (positive value, or red colour in Figures~\ref{fig:1cav_KTOatSides_Efield_Zoom_er3000} and \ref{fig:1cav_KTOatSides_Efield_Zoom_er5000}) is more concentrated at the ferroelectric material. This small change in the electric field pattern of the mode provides the required resonant frequency shift leading to the change in the axion search frequency.\\

\section{The KTO ferroelectric as an interresonator coupling element}
\label{KTOasCoupler}
Next we will see how ferroelectric KTO films can be used instead of iris windows (Figure~\ref{fig:IrisCouplings}) as interresonator coupling elements between adjacent cavities in a multicavity haloscope. The iris windows often cause problems in the quality factor when manufacturing due to fabrication tolerances and cuts in regions with high surface current levels, leading in some cases to misalignment issues and low $Q_0$s. With the use of ferroelectric films only one cut (in a plane with a low surface current level) would be necessary for manufacturing the housing, introducing narrow slits to place the dielectric elements at the proper position inside the metallic box.
\subsection{Modelling}
\label{KTOfilmAsCoupling}
\begin{figure}[h]
\centering
\includegraphics[width=0.5\textwidth]{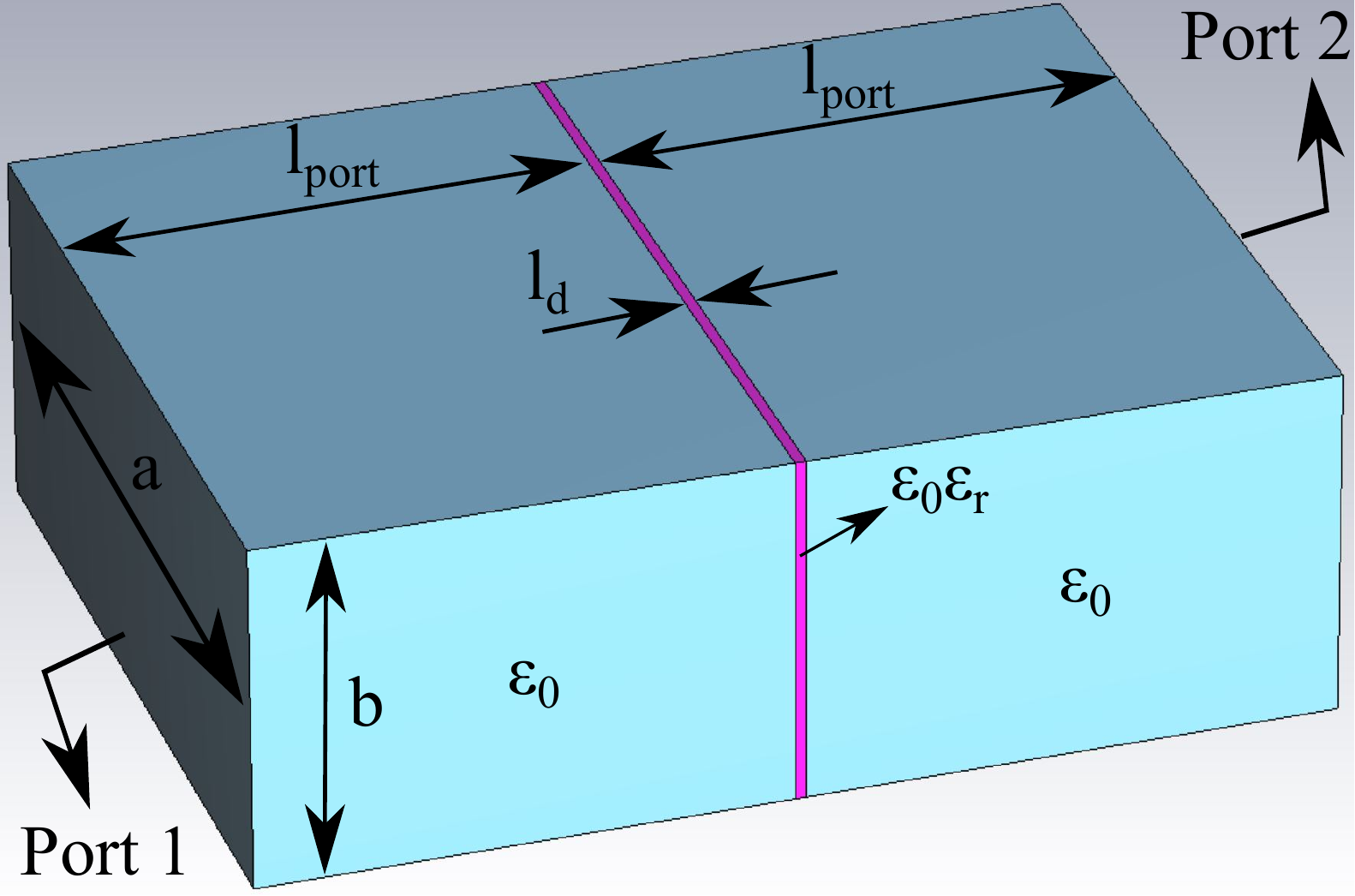}
\caption{\label{fig:KTOasCoupling} 
Characterization of a ferroelectric KTO as a coupling element. The system is based on three waveguide sections: the first and last ones are hollow waveguides ($\varepsilon = \varepsilon_0$), and the second one corresponds to the KTO ($\varepsilon = \varepsilon_0 \, \varepsilon_r$) filling homogeneously the height and the width dimensions. The waveguide dimensions are those of the WR-90 standard rectangular waveguide. The length of the KTO ($l_d$) and its relative permittivity ($\varepsilon_r$) control the coupling. The variable $l_{port}$ is the distance of the KTO to the reference plane of both ports.}
\end{figure}

Figure~\ref{fig:KTOasCoupling} shows the structure that is used to examine the effect of KTO ferroelectrics as interresonator couplings in rectangular waveguides. It consists of two waveguide sections of length $l_{port}$ connected by a KTO film of thickness $l_d$. Waveguide ports are used here instead of coaxial ports in order to simplify the modelling of the structure.\\

\begin{figure}
\centering
\includegraphics[width=0.8 \textwidth]{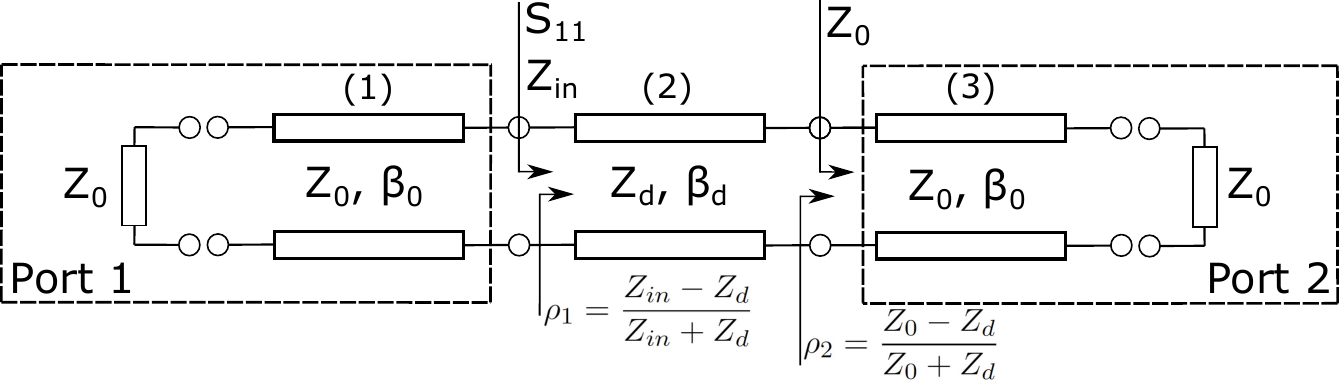}

\vspace{0.5cm}
\includegraphics[width=0.8 \textwidth]{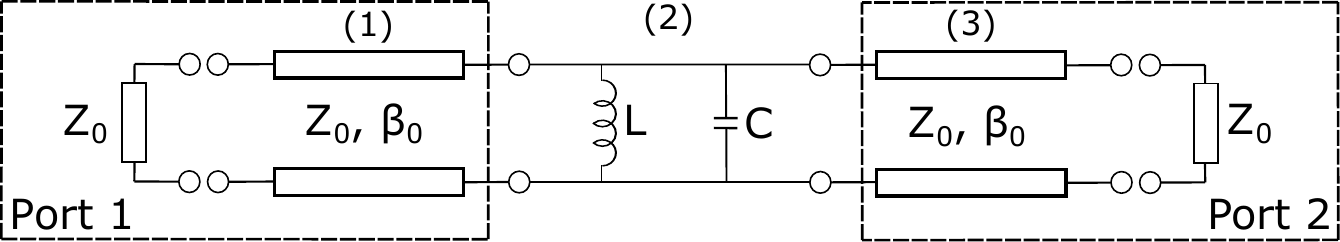}

\vspace{1cm}
\includegraphics[width=0.8 \textwidth]{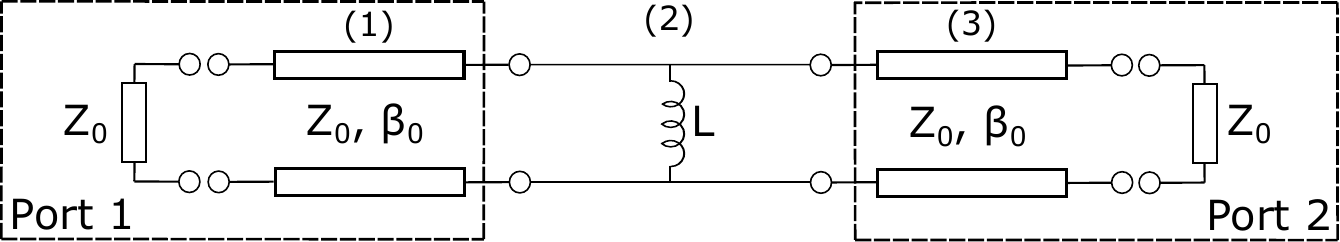}

\vspace{1cm}
\includegraphics[width=0.8 \textwidth]{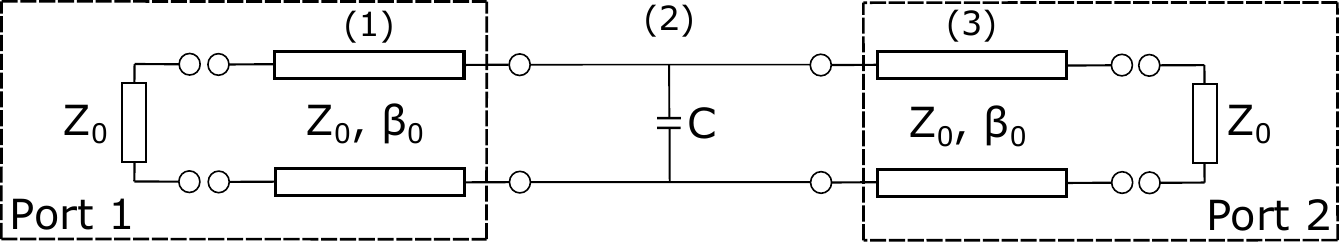}
\caption{\label{fig:LineKTOLine_Circuit} 
Analysis of the system with transmission lines using a single mode representation. From top to bottom: three transmission lines, KTO as LC resonator, KTO as inductive coupling ($f<f_{KTO}$), and KTO as capacitive coupling ($f>f_{KTO}$).}
\end{figure}
In order to analyze this system we can develop a simple single-mode transmission line model (see first circuit of Figure~\ref{fig:LineKTOLine_Circuit}). The hollow waveguide regions are represented by the first (1) and the last (3) transmission lines filled with vacuum ($\varepsilon=\varepsilon_0$), while the KTO film is depicted by the central transmission line, representing an homogeneous medium with permittivity $\varepsilon=\varepsilon_0 \, \varepsilon_r$. The characteristic impedances for the three lines ($Z_0$ and $Z_d$) are the modal impedances of the fundamental mode ($TE_{10}$) in each region, given by \cite{pozar}:
\begin{equation}
\label{eq:Zci}
Z_{i}=Z_{i,TE_{10}}=\frac{2 \, \pi \, f \,  \mu_0}{\beta_i}=\frac{\eta/\sqrt{\varepsilon_{ri}}}{\sqrt{1-(f_{ci} \, / \, f_0)^2}},
\end{equation}
where $\beta_i$ is the propagation constant of the fundamental mode ($TE_{10}$) giving by $\beta_i = (2\pi)/\lambda_{gi}$ (being $i = d$ for line (2) and $i = 0$ for lines (1) and (3)), $f_{ci} = c/(2 \, a\sqrt{\varepsilon_{ri}})$ is the cut-off frequency of the $TE_{10}$ mode in each waveguide region, and $\eta = \sqrt{\mu_0 / \varepsilon_0} \, \simeq \, 120\pi \, \Omega$ the free space impedance.\\

Due to the high relative permittivity change between the KTO (medium (2)) and the hollow waveguide (media (1) and (3)), a high impedance step between ferroelectric and vacuum appears. In this situation, the central transmission line behaves as a resonator when its length is close to half guided wavelength ($l_d=\lambda_{gKTO}/2$) \cite{pozar}. This situation can be represented with a parallel lumped LC resonator, as shown in the second circuit of Figure~\ref{fig:LineKTOLine_Circuit}, which resonates at frequency $f_{KTO}=\frac{1}{2\pi\sqrt{LC}}$, where $L$ and $C$ are the inductance and capacitance values, respectively.\\

For $f<f_{KTO}$, the admitance\footnote{The admitance is the inverse of the impedance ($Y=1/Z$). The use of this parameter is usually employed in the analysis of parallel circuits.} of the inductor ($Y_L=\frac{1}{j2\pi f L}$) is larger than the admitance of the capacitor ($Y_C=j2\pi f  C$), leading to an inductive behaviour. In fact, for these frequencies the current of the circuit circulates mainly through the inductive component. Therefore, the KTO acts as an inductive coupling, as shown in the third circuit of Figure~\ref{fig:LineKTOLine_Circuit}. Analogously, for $f>f_{KTO}$ the KTO acts as a capacitive coupling (see fourth circuit in Figure~\ref{fig:LineKTOLine_Circuit}). On the other hand, note that the relative permittivity of the KTO can be used to easily change its resonant frequency $f_{KTO}$. This can be used to conveniently adjust the frequency regions where the KTO behaves as inductive or as capacitive coupling. This will be elaborated further in the next subsection, when we design the KTO as coupling element introducing some examples.\\

Taking the model of Figure~\ref{fig:KTOasCoupling}, we have characterized its behaviour by analyzing the first circuit of Figure~\ref{fig:LineKTOLine_Circuit}, from where the scattering parameters can be extracted with the following equations:
\begin{equation}
\label{eq:S11}
S_{11}=\frac{Z_{in}-Z_{0}}{Z_{in}+Z_{0}},
\end{equation}

\begin{equation}
\label{eq:S21}
S_{21}=\frac{(1+S_{11})(1+\rho_2)e^{- \, j \,\beta_d \, l_d}}{1+\rho_1},
\end{equation}
where $S_{11}$ and $S_{21}$ are the reflection and transmission parameters, respectively, $Z_{in}$ is the input impedance, $Z_{0}$ is the characteristic impedance of lines (1) and (3), $\rho_2$ is the reflection coefficient of line (3) referred to line (2), $j \equiv \sqrt{-1}$ is the complex imaginary unit, $\beta_d$ is the longitudinal component of the propagation vector in medium (2), given by $\beta_d=(2\pi)/\lambda_{gd}$, and $\rho_1$ is the reflection coefficient at the input referred to line (2). $\lambda_{gd}$ can be computed with equation \ref{eq:ld} employing the KTO cut-off frequency as $f_{c}^{KTO} = c/(2a\sqrt{\varepsilon_r})$. $Z_{in}$ is calculated using transmission line theory as follows:
\begin{equation}
\label{eq:Zin}
Z_{in}=Z_{d} \, \frac{Z_{0} + j \, Z_{d} \, \tan{(\beta_d \, l_d)}}{Z_{d} + j \, Z_{0} \, \tan{(\beta_d \, l_d)}}.
\end{equation}\\

\subsection{Results for the KTO as coupling element}
\label{KTOasCouplingResults}
\begin{figure}[h]
\centering
\includegraphics[width=0.8\textwidth]{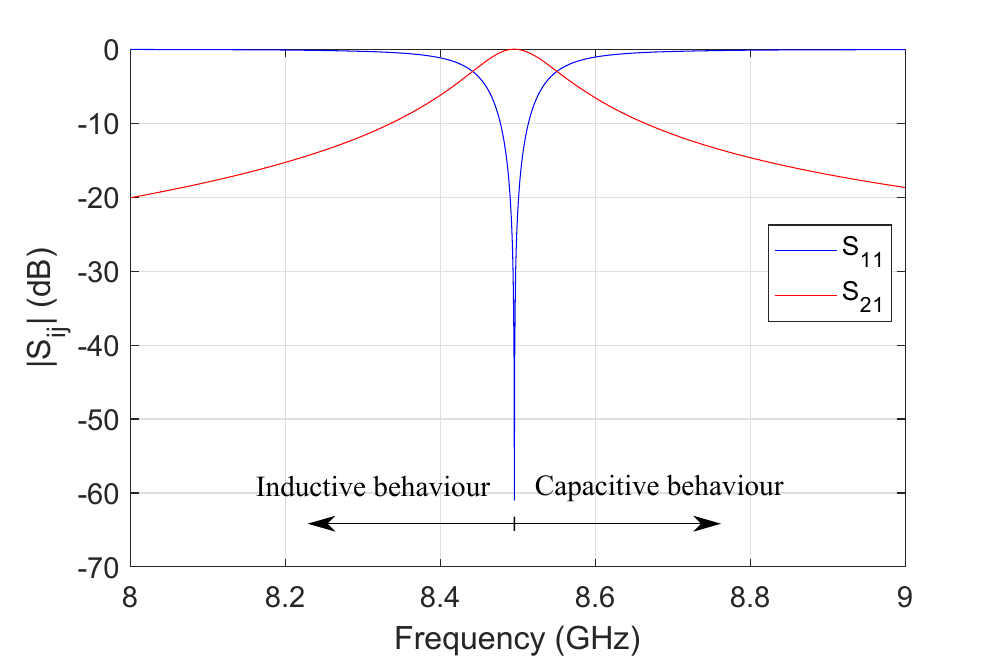}
\caption{\label{fig:2TL&KTOasCoupling_pS} 
Magnitude of the scattering parameters from the model of Figure~\ref{fig:KTOasCoupling} using equations \ref{eq:S11} and \ref{eq:S21}. The KTO film behaves as inductive coupling in the region $f<f_{KTO}$ while it acts as capacitive coupling in the region $f>f_{KTO}$.}
\end{figure}

Using the equations from subsection~\ref{KTOfilmAsCoupling} and taking as example $\varepsilon_r=4000$ and $l_d=\lambda_g/2=279\, \mu$m for a KTO resonance at $f_{KTO}=8.5$~GHz, the results shown in Figure~\ref{fig:2TL&KTOasCoupling_pS} are obtained.
As it was discussed in the previous subsection, for $f<f_{KTO}=8.5$~GHz, the coupling is inductive and for $f>f_{KTO}=8.5$~GHz, the coupling is capacitive. It is evident that by adjusting the KTO resonant frequency $f_{KTO}$, the frequency regions where the film acts as inductive or capacitive coupling can be easily tuned. The adjustment of the KTO resonant frequency can be easily achieved by changing its thickness ($l_d$) or/and the relative permittivity ($\varepsilon_r$).\\

\begin{figure}[h]
\centering
\includegraphics[width=0.75\textwidth]{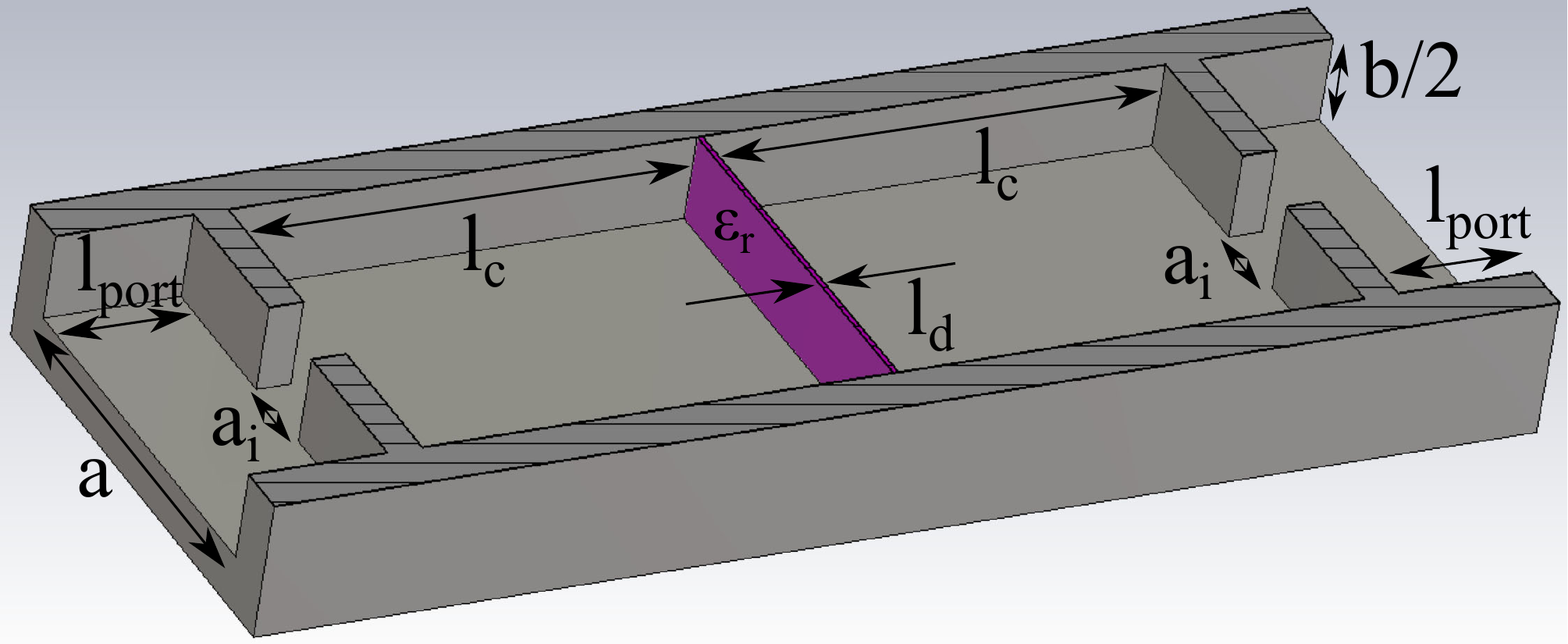}
\caption{\label{fig:2cav_KTOiris_WGports} 
Model based on two resonant cavities connected by an interresonator coupling implemented by the dielectric KTO film. The input/output couplings are implemented with standard inductive windows. The picture shows the symmetric half, being the dashed region the symmetry plane. The dimensions are the same as the model from Figure~\ref{fig:KTOasCoupling}, but using $l_{port}$ of 8~mm. The length of the cavities $l_c$ is set for a desired working frequency $f_0=8.42$~GHz, while the thickness of the KTO $l_d$ is adjusted to have its resonant frequency $f_{KTO}$ above or below the working frequency in order to implement either inductive or capacitive coupling. The input/output iris width is fixed small ($a_i = 5$~mm) for reducing the load effect.}
\end{figure}

For the RADES designs the extraction of the so-called physical coupling $k$ value is usually needed \cite{RADES_paper1}. This value is obtained with a model based on two resonant cavities connected with one interresonator coupling, which will be implemented in this case with the dielectric KTO slab. We have employed the physical model of Figure~\ref{fig:2cav_KTOiris_WGports}, where inductive windows have been selected to implement the input/output couplings, and the KTO film acts as interresonator coupling.
Using this structure, the $k$ value can be obtained as \cite{Cameron} (subsection $14.2.2$):
\begin{equation}
\label{eq:k}
k=\frac{f_{even}^2-f_{odd}^2}{f_{even}^2+f_{odd}^2},
\end{equation}
where, $f_{even}$ and $f_{odd}$ are the even (or magnetic) and odd (or electric) frequencies, respectively \cite{Cameron}.\\

\begin{figure}[h]
\centering
\begin{subfigure}[b]{0.85\textwidth}
         \centering
         \includegraphics[width=1\textwidth]{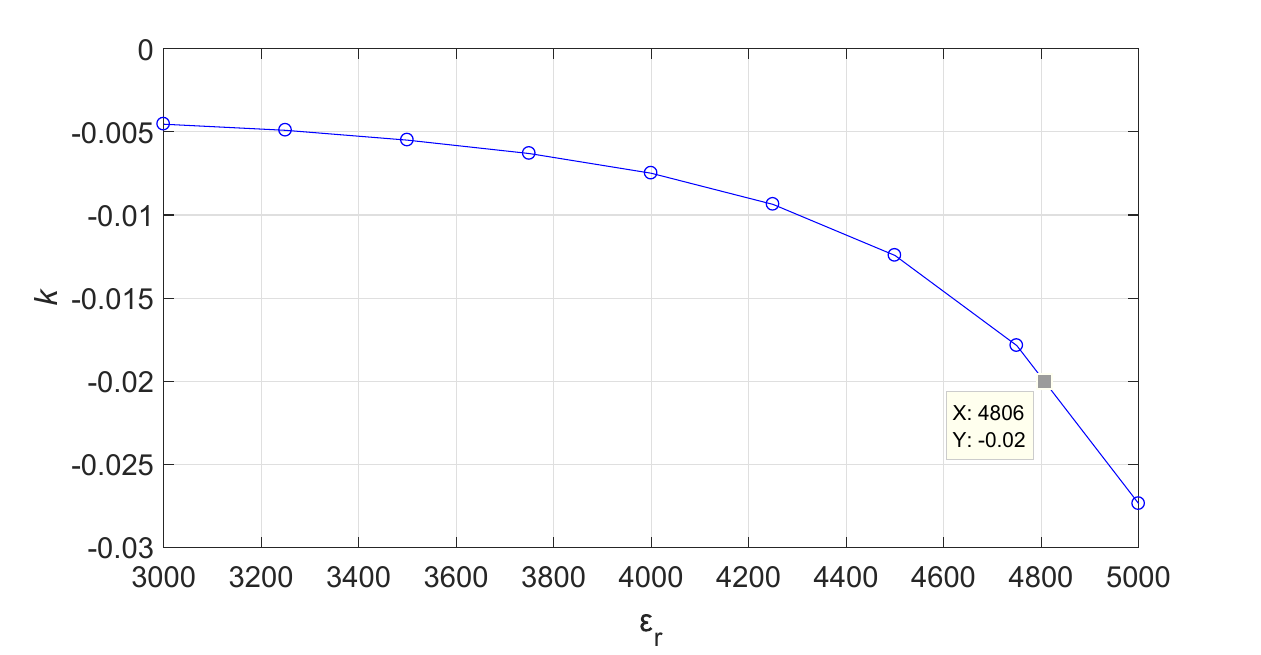}
         \caption{}
         \label{fig:kIND}
\end{subfigure}
\hfill
\begin{subfigure}[b]{0.85\textwidth}
         \centering
         \includegraphics[width=1\textwidth]{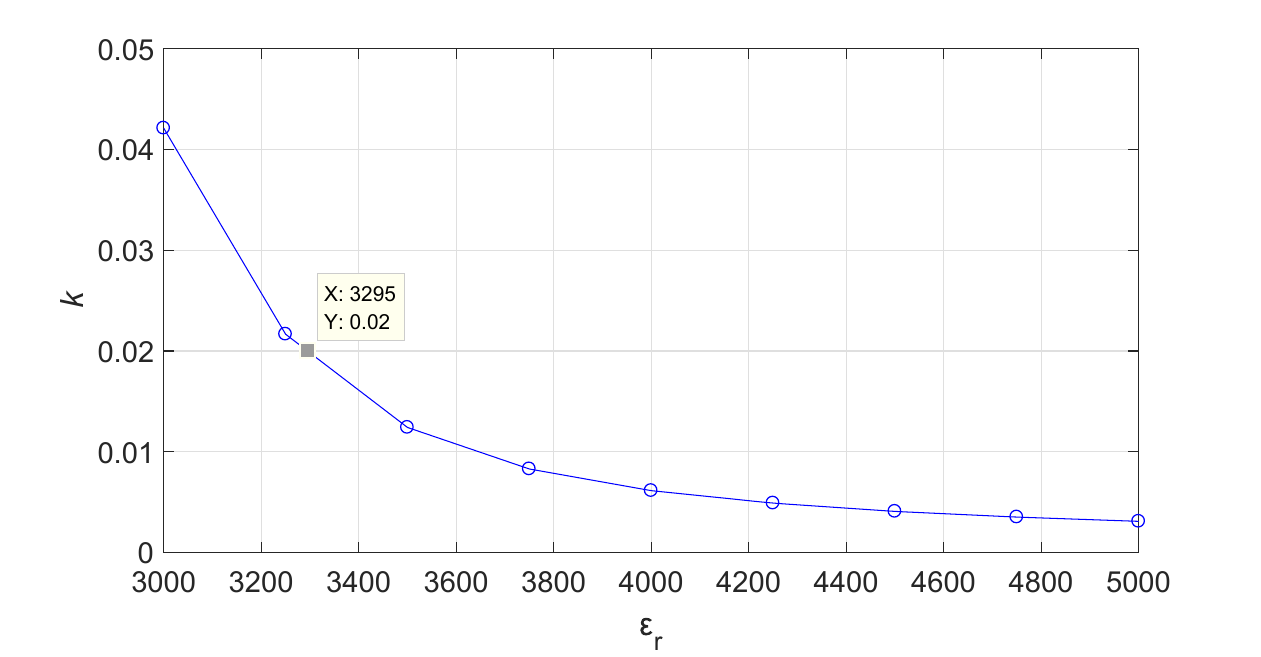}
         \caption{}
         \label{fig:kCAP}
\end{subfigure}
\caption{Physical coupling value ($k$) versus KTO relative permittivity for (a) inductive behaviour ($l_d = 250\, \mu$m) and (b) capacitive behaviour ($l_d = 316\, \mu$m). In both cases coupling values are calculated at the working frequency of $f_0=8.42$~GHz. Simulation results obtained from CST \cite{CST}.}
\label{fig:k_vs_EpsR}
\end{figure}

Following this procedure, the next step is to characterize the KTO as a coupling element at a working frequency of $f_0=8.42$~GHz. First, we will extract the KTO inductive coupling. For this case, the KTO resonance is adjusted for a frequency higher than $8.42$~GHz.  In our test we have selected a KTO resonant frequency of $f_{KTO} = 9.5$~GHz, which corresponds with a length $l_d = 250\, \mu$m for a relative permittivity of $\varepsilon_r = 4000$. We assume a KTO dielectric constant range $\varepsilon_r \in [3000 , 5000]$. In a second step, for the capacitive coupling, we select $l_d=316\, \mu$m, which leads to a KTO resonant frequency of $f_{KTO}=7.5$~GHz for $\varepsilon_r = 4000$, lower than the target frequency of $f_0=8.42$~GHz. Figure~\ref{fig:k_vs_EpsR} plots the values of the calculated couplings as a function of the permittivity range of the KTO for both inductive and capacitive cases. Results demonstrate that a ferroelectric film is indeed able to act as a coupling element, providing coupling values in a range from close to zero to maximum values around $|k|=0.04$, and with both signs (negative for inductive and positive for capacitive).\\

A typical realizable value of $|k|$ for metallic irises is $0.02$ \cite{RADES_paper2}. To prove the effectiveness of the KTO as coupling, it is shown how this value can be obtained with a ferroelectric film. As it can be observed in Figure~\ref{fig:k_vs_EpsR}, the $|k|=0.02$ value is obtained at $\varepsilon_r=4806$ for the inductive case, and at $\varepsilon_r=3295$ for the capacitive case.\\

In this way, the RADES alternating coupling structures \cite{RADES_paper2} can be designed by alternating low and high permittivities, which will give to each coupling its necessary properties. As discussed in that paper, alternating coupling increases the mode separation. In addition, this system can provide a required final adjustment for the couplings in case of manufacturing errors or if the response needs to be modified for any reason. This can simply be done by adjusting the KTO permittivity values using the biasing mechanisms discussed in the next section.\\

\begin{figure}[h]
\centering
\begin{subfigure}[b]{0.7\textwidth}
         \centering
         \includegraphics[width=1\textwidth]{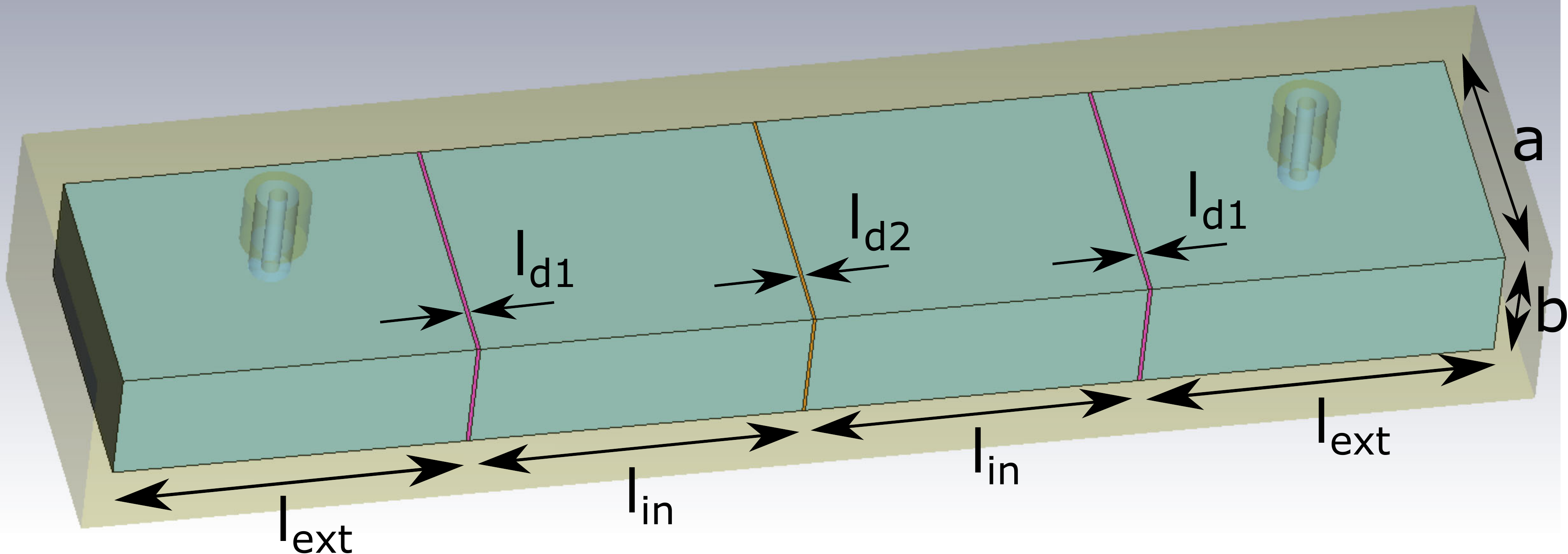}
         \caption{}
         \label{fig:4cav_KTOasCouplings_Model}
\end{subfigure}
\hfill
\begin{subfigure}[b]{0.9\textwidth}
         \centering
         \includegraphics[width=1\textwidth]{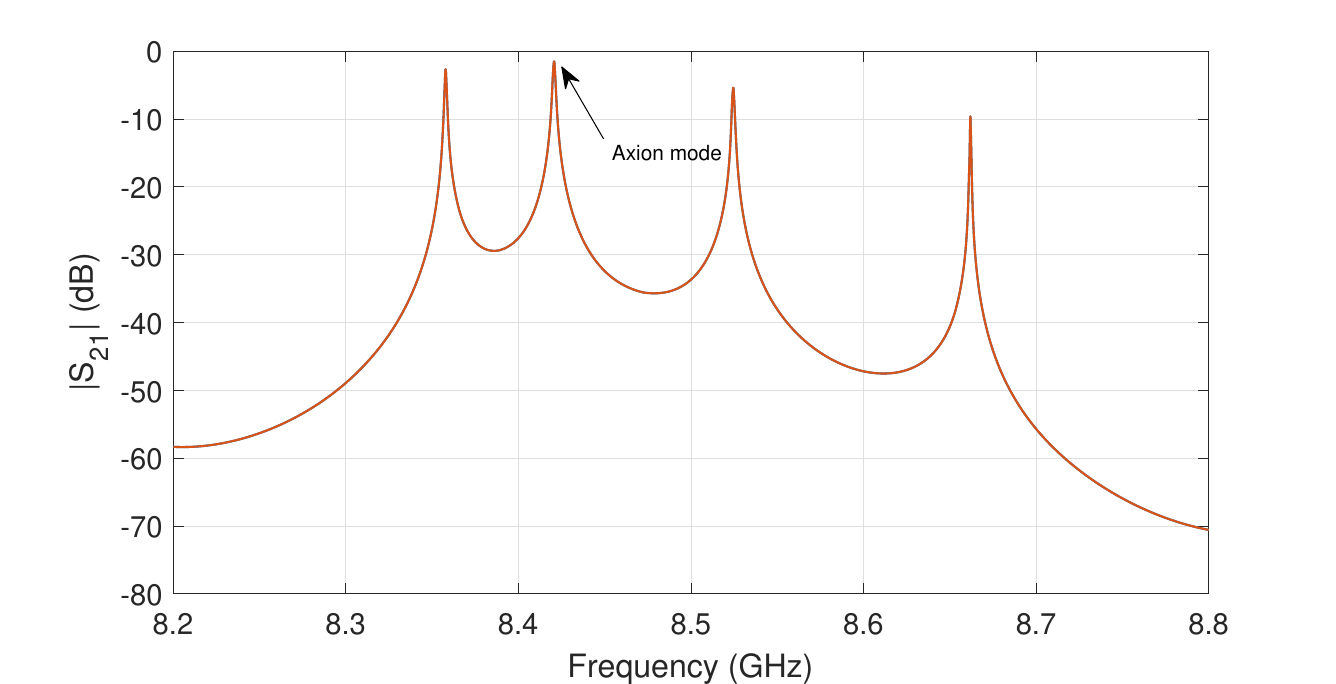}
         \caption{}
         \label{fig:4cav_KTOasCouplings_Sp}
\end{subfigure}
\hfill
\begin{subfigure}[b]{0.9\textwidth}
         \centering
         \includegraphics[width=1\textwidth]{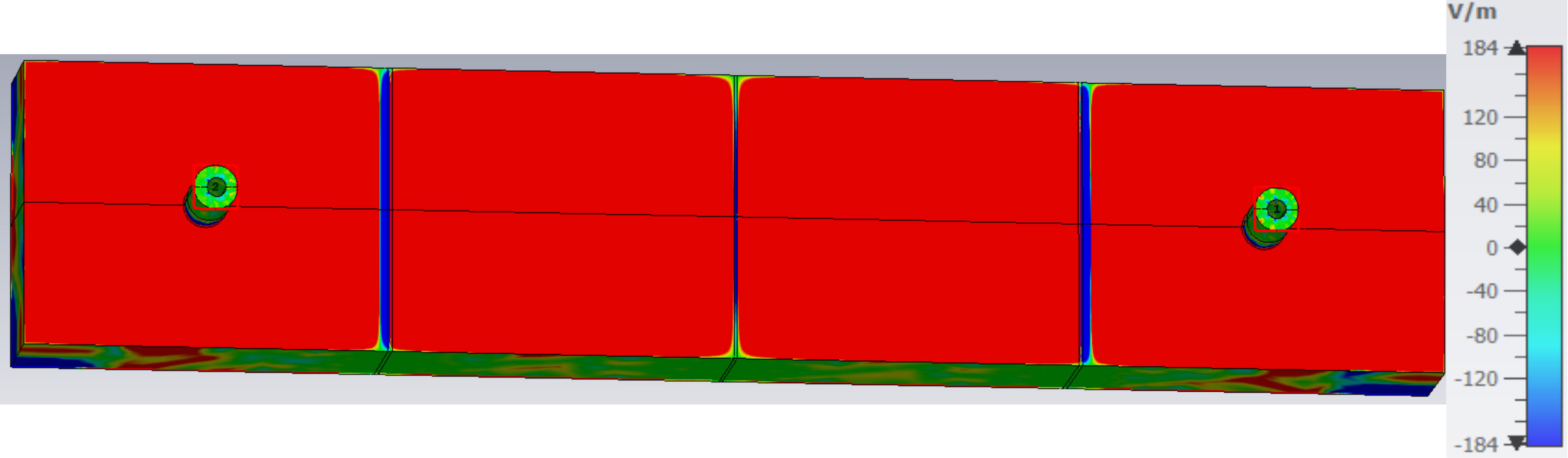}
         \caption{}
         \label{fig:4cav_KTOasCouplings_Efield}
\end{subfigure}
\caption{(a) Model based on a four-subcavities haloscope with coaxial input/output ports. The length of the two end cavities is $l_{ext}=29.05$~mm, the length of the two inner cavities is $l_{in}=27.34$~mm. (b) Transmission $S_{21}$ parameter magnitude. (c) Electric field (vertical component) of the axion mode (at $f_0=8.42$~GHz). Similarly to Figure~\ref{fig:4cav_KTOatSides_Efield} a strong colour scaling has been applied to correctly appreciate the electric field levels. Simulation results obtained from CST \cite{CST}.}
\label{fig:4cav_KTOasCouplingsModel&Sp&Efield}
\end{figure}

The previous example can be used to conceive a more practical structure with the design of four-subcavities connected with three alternating KTO couplings (two of them capacitive and one inductive) and coaxial input/output ports, as shown in Figure~\ref{fig:4cav_KTOasCouplings_Model}. This method provides an improvement in the mode separation parameter with respect to an all-inductive or all-capacitive design \cite{RADES_paper2}. Employing the previous coupling value ($|k|=0.02$) for all the couplings, we have to use the obtained KTO parameters, $l_d = 316\, \mu$m, $\varepsilon_r=3295$ for the capacitive coupling (pink colour films in Figure~\ref{fig:4cav_KTOasCouplings_Model}), and $l_d = 250\, \mu$m, $\varepsilon_r=4806$ for the inductive coupling (brown colour film in Figure~\ref{fig:4cav_KTOasCouplings_Model}). The coaxial probes are optimized for a weak input/output coupling. Figure~\ref{fig:4cav_KTOasCouplings_Sp} shows the transmission response obtained with these KTO films. Due to the alternating behaviour in the interresonator couplings, the axion mode corresponds now to the second resonance \cite{RADES_paper2}. Also, Figure~\ref{fig:4cav_KTOasCouplings_Efield} shows the electric field of the axion mode, where the negative electric field value can be observed inside the KTOs (blue color), as expected. It is interesting to note that the design has been very effective since the field is perfectly synchronous in all the cavities as it is required to maximize the axion form factor \cite{RADES_paper2}. The quality and form factor obtained for this mode are $Q_{0}=31668$ and $C=0.586$, and the resulting mode separation ($63$~MHz or $0.75\, \%$) is very good (see Figure~\ref{fig:4cav_KTOasCouplings_Sp}), demonstrating that this concept can be used for the design of real haloscopes with the added flexibility of allowing an easy combination of inductive and capacitive couplings.\\

Finally, the use of ferroelectrics as interresonator coupling elements avoids the need to manufacture irises based on windows (as in Figure~\ref{fig:IrisCouplings}) which have caused many alignment and contact problems for the haloscopes in previous RADES implementations with subsequent degradation in the quality factors.

\section{Electrical biasing of ferroelectrics}
\label{Biasing}
The ferroelectric permittivity can be changed by applying a static voltage or by modifying the operation temperature. For the voltage option, an asynchronous system can be designed where each ferroelectric film can be tuned individually. On the other hand, for the temperature option the whole haloscope will be cooled/warmed, so all the ferroelectrics would have the same $\varepsilon_r$ value. In the last option, the implementation of an alternating coupling system (inductive + capacitive couplings in the same structure) with ferroelectrics needs to take into account some details. A possibility is to have longer thickness $l_d$ for the ferroelectrics that will behave as capacitive couplings, adjusting the frequency $f_{KTO}$ below the operating frequency (see Figure~\ref{fig:2TL&KTOasCoupling_pS}). In this way the coupling value would change if the permittivity decreases, but ensuring always that it operates inside the capacitive region. Similarly, for the ferroelectrics operating as inductive couplings, the thickness $l_d$ will be shorter to ensure that the KTO resonant frequency $f_{KTO}$ is quite high in frequency to ensure that the coupling value would change in the region where it behaves as inductive coupling (see Figure~\ref{fig:2TL&KTOasCoupling_pS}). If  the KTO is controlled with the voltage technique, the thickness $l_d$ could be the same for all the coupling ferroelectrics and the coupling behavior will be adjusted by applying different voltages to each slab.\\

Since the haloscope should operate at the minimum temperature, an increment in temperature to control de KTO films will reduce the sensitivity of the experiment, since $T_{sys}$ will increase and $Q_0$ will decrease. Therefore, the best option to change the permittivity of the KTO films appears to be the electrical biasing.\\

\begin{figure}[h]
\centering
\includegraphics[width=0.6\textwidth]{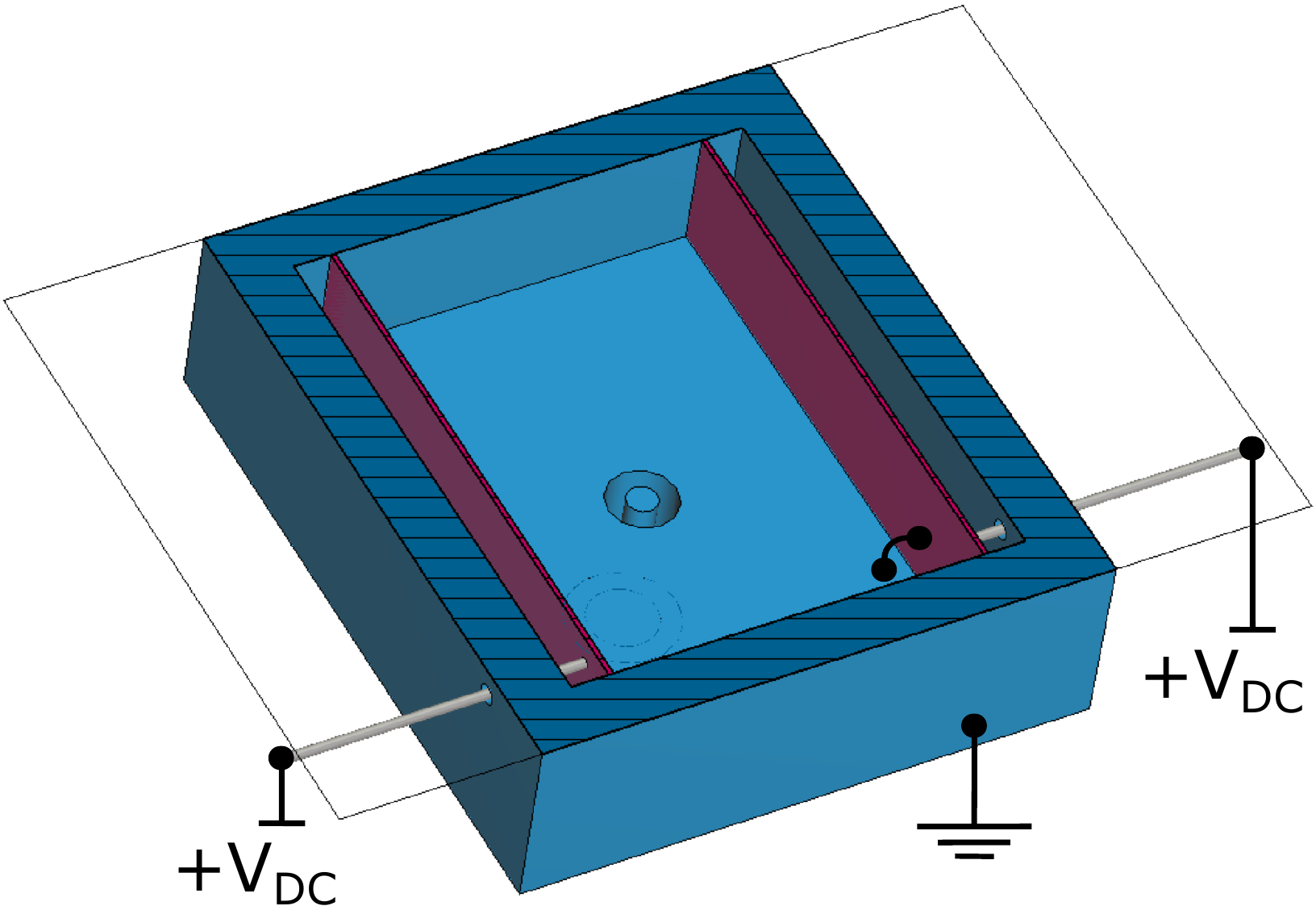}
\caption{\label{fig:Electrodes} Horizontal cut of the resonant cavity (blue material) of Figure~\ref{fig:1cav_KTOatSides_Model} showing the position of the electrodes for the ferroelectric biasing. Ferroelectric KTOs are depicted in pink colour. $V_{DC}$ represents the DC voltage applied to the electrodes. One of the two electrodes in each KTO is connected to the cavity housing which acts as ground.}
\end{figure}

In the application of a static voltage to these films, the position of the electrodes has to be considered carefully. Small holes at the haloscope walls are needed to provide internal access for the biasing DC cables. In addition, a suitable notch filter should be designed at the biasing section for avoiding high losses (and $Q_0$ reduction) due to the parasitic elements that could be introduced from non desirable frequencies. The electrodes would be positioned at the two closest surfaces, as it is shown in Figure~\ref{fig:Electrodes}. The biasing cables and access openings are positioned close to the corners where the surface current is minimum for the $TE_{101}$ mode.\\

DC bias contacts and deposition technology strongly depend on the haloscope cavity design \cite{EuclidTechlabs}. If the operational mode of the design has no electric field components parallel to the KTO surface, the DC contacts can be made either with high conductivity or with superconducting materials. In the case of normal conducting electrodes, high conductivity materials can be deposited and adhesion/thermal expansion matching layers are applied to ensure low microwave losses in the electrodes. An alternative option can be superconducting contact deposition. In \cite{Spartak}, a resonator was fabricated on single-crystal (100) KTO discs $0.5\times10$~mm and epitaxial-grade polished. Double-sided superconducting YBCO (or Yttrium Barium Copper Oxide) films were deposited using the co-evaporation technique \cite{Spartak}.\\

In many currently considered haloscope cavity designs such as the case studied in this paper, the operating TE modes have their field components parallel to the DC contact surface for the KTO films. In this case, to prevent $Q_0$ degradation high resistivity contacts can be used to introduce the biasing. In this scenario the conducting but high resistivity DC bias contacts have to be deposited on the crystal surface, as discussed in \cite{EuclidTechlabs}.\\

Finally, it is essential that the voltage supply does not generate noise at the frequency range of the cavity in order to avoid the injection of an additional noise source. This noise can be reduced to a minimum value with careful design and low-pass filtering. It is also important to guarantee a correct shielding to the DC voltage cables for avoiding the introduction of pick-up noise.

\section{Conclusions and prospects}
\label{Conclusions}
This paper proposes for the first time the possibility of employing electronic tuning systems based on ferroelectric materials. Design guidelines and very interesting simulation results are provided for the design of a tunable haloscope sweeping a considerable spectral region of axion masses. A haloscope based on four subcavities and three inductive interresonator couplings (iris windows) has been designed with a good tuning range and right results in the quality and form factors.\\

Also, in this work the implementation of any type of interresonator coupling (inductive or capacitive) with the use of KTO ferroelectric films is demonstrated. A haloscope based on four subcavities and three interresonator ferroelectric alternating couplings has been designed with good form factor and unloaded quality factor results. The structure employs an alternating behaviour in the couplings: two capacitives and one inductive in order to improve separation of neighbour modes. This eliminates the need to manufacture metallic iris windows and could improve the quality factor of the structures. This work demonstrates that the concept allows to combine very easily interresonator couplings of inductive and capacitive types in the same structure.\\

Two practical multicavity haloscopes incorporating ferroelectric films for frequency and coupling tuning have been designed to demonstrate the proposed concepts. The main future research line of this work is the measurement of these KTO films in real prototypes implementing an adequate biasing system with the minimum impact on the form factor and the quality factor of the axion mode. Also, the combination of both ideas (tuning and coupling with ferroelectrics) in a single prototype is being investigated.

\acknowledgments
This work was performed within the RADES group; we thank our colleagues for their support. In addition, this work has been funded by the grant PID2019-108122GB-C33, funded by MCIN/AEI/10.13039/501100011033/ and by "ERDF A way of making Europe". JMGB thanks the grant FPI BES-2017-079787, funded by MCIN/AEI/10.13039/501100011033 and by "ESF Investing in your future". Also, this project has received partial funding through the European Research Council under grant ERC-2018-StG-802836 (AxScale).

\bibliographystyle{JHEP.bst}
\bibliography{mybibfile.bib}
\end{document}